\documentclass[pdflatex,sn-mathphys-num]{sn-jnl}

\usepackage{graphicx}%
\usepackage{multirow}%
\usepackage{amsmath,amssymb,amsfonts}%
\usepackage{amsthm}%
\usepackage{mathrsfs}%
\usepackage[title]{appendix}%
\usepackage{xcolor}%
\usepackage{textcomp}%
\usepackage{manyfoot}%
\usepackage{booktabs}%
\usepackage{algorithm}%
\usepackage{algorithmicx}%
\usepackage{algpseudocode}%
\usepackage{listings}%

\theoremstyle{thmstyleone}%
\theoremstyle{thmstyletwo}%

\theoremstyle{thmstylethree}%

\begin{document}

\title[Onset of three-dimensionality in transonic shock–boundary-layer interaction over a natural laminar flow airfoil]{Onset of three-dimensionality in transonic shock–boundary-layer interaction over a natural laminar flow airfoil}


\author[1]{\fnm{Abhijeet} \sur{Guha}}
\equalcont{These authors contributed equally to this work.}

\author*[1]{\fnm{Aditi} \sur{Sengupta}}\email{aditi@iitism.ac.in}
\equalcont{These authors contributed equally to this work.}

\affil[1]{\orgdiv{Department of Mechanical Engineering}, \orgname{Indian Institute of Technology}, \city{Dhanbad}, \postcode{826004}, \state{Jharkhand}, \country{India}}


\abstract{The development of three-dimensionality in transonic shock--boundary-layer interactions (SBLI) can significantly influence the aerodynamic response of natural laminar flow (NLF) airfoils, yet the associated vorticity dynamics remain insufficiently understood. The present study investigates the short-time evolution of transonic SBLI over the SHM1 NLF airfoil using high-fidelity three-dimensional (3D) simulations at three operating conditions reported by Fujino {\it et al.} [“Natural-laminar-flow airfoil development for a lightweight business jet,” J. Aircraft 40, 609-615 (2003)]. The selected cases represent three distinct operating regimes: the initial development of a normal shock, drag divergence, and shock-induced separation. Comparisons with previously reported two-dimensional (2D) simulations establish the temporal interval over which the spanwise-averaged 3D flow remains close to the corresponding 2D evolution, while departures in surface pressure and skin-friction responses identify the subsequent development of 3D effects within the SBLI region. The temporal evolution of volume-integrated streamwise and wall-normal enstrophies further characterizes the development of the associated 3D vortical field. The compressible enstrophy transport equation (CETE) is subsequently employed to examine the mechanisms governing the evolution of the 3D vortical field. Viscous and compressibility-related terms dominate the normalized enstrophy budget during the early evolution, while vortex stretching remains a minor contribution in the normal-shock and drag-divergence regimes but becomes appreciable, albeit intermittently, with the development of shock-induced separation. Across the three operating regimes, the same principal enstrophy-transport mechanisms remain active, but their relative contributions to the normalized budget are redistributed with the SBLI topology.}




\maketitle

\section{Introduction}\label{sec1}

The aviation industry continues to pursue technologies that improve aerodynamic efficiency while reducing fuel consumption and greenhouse gas emissions. Among the various approaches proposed, natural laminar flow (NLF) airfoils have emerged as an effective passive strategy for reducing skin-friction drag by delaying the transition from laminar to turbulent boundary layers \citep{krishnan2017review, karpuk2024investigation}. Since viscous drag constitutes a substantial fraction of the total drag during cruise conditions, extending laminar flow over a larger portion of the airfoil can significantly improve fuel efficiency. Consequently, NLF technology has evolved from a conceptual design philosophy to an integral component of modern business aircraft and next-generation transport configurations, including the SHM1 airfoil developed for the Honda Jet \citep{fujino2003natural}. Advances in high-fidelity numerical simulations have further accelerated the development of NLF configurations by enabling accurate prediction of laminar boundary-layer evolution, shock formation, and aerodynamic performance under realistic operating conditions \citep{halila2019effects, sengupta2021thermal}.

Although NLF airfoils are designed to maximize laminar flow under cruise conditions, they frequently encounter off-design operating environments where localized supersonic regions terminate in shock waves. The interaction of these shocks with the boundary layer produces complex transonic shock boundary-layer interactions (SBLI), characterized by strong adverse pressure gradients, flow deceleration, and, under sufficiently strong interactions, shock-induced separation \cite{dolling2001fifty, babinsky2011shock}. These phenomena modify the pressure distribution, increase aerodynamic drag, and alter the stability of the boundary layer, thereby reducing the aerodynamic benefits of laminar flow \citep{cole2012transonic, sengupta2022comparative}. Consequently, understanding the evolution of transonic SBLI is essential for accurately predicting the aerodynamic performance and operational envelope of NLF airfoils.

Over the past several decades, considerable effort has been devoted to understanding transonic SBLI through experimental investigations, high-fidelity numerical simulations, and theoretical analyses. Experimental studies have documented the evolution of shock structures, separation bubbles, and unsteady pressure fields in canonical SBLI configurations \citep{barre1996experimental, dussauge2008shock, bruce2010experimental}. Direct numerical simulations (DNS) and large-eddy simulations (LES) have further revealed the intricate coupling between shock motion, boundary-layer development, and vortical structures, providing valuable insight into the dynamics of separated compressible flows \citep{larsson2013reynolds, toure2018numerical, gross2018numerical, quadros2018numerical}. Such simulations have also demonstrated the sensitivity of the resolved flow structures to the numerical resolution, particularly in regions downstream of the shock interaction where characteristic flow scales can become small \citep{ceci2025grid}. For NLF airfoils, high-fidelity simulations have similarly been used to examine the development of shock-induced separation and its influence on the aerodynamic response under off-design conditions \citep{sengupta2024thermal}. An additional aspect of SBLI concerns its inherently 3D nature. Reviews of 3D SBLI have shown that spanwise flow, secondary motions, and pressure-wave interactions can substantially modify the separation topology and associated vortical structures relative to nominally 2D interactions \citep{gaitonde2023dynamics, sabnis2023review}. The development of three-dimensionality is widely recognized as an inherent characteristic of sufficiently strong transonic SBLI. As the interaction strengthens, spanwise variations emerge within the separated shear layer, shock foot, and wake, progressively modifying both the near-wall flow and the downstream vortical structures \cite{tijdeman1977investigations}. High-fidelity simulations of transonic airfoils have further revealed complex interactions among shock motion, separated shear layers, acoustic disturbances, and 3D vortical structures \cite{zauner2019direct}. The distinction between 2D and 3D interactions is particularly important in separated flows, for which the organization of vortical structures and the spatial distribution of the separated region can differ substantially from those obtained from 2D representations \citep{gaitonde2023dynamics}. These observations establish the importance of 3D effects in the fully developed interaction; however, they do not by themselves describe how such effects emerge during the initial evolution of an interaction that is initially close to 2D. 

Addressing this question requires a framework capable of directly quantifying the mechanisms responsible for vorticity generation and evolution in compressible flows. While vorticity transport equations provide valuable insight for incompressible flows, compressible SBLI introduces additional physical processes arising from density variations, dilatation, and baroclinic torque \cite{sengupta2024thermal}. The compressible enstrophy transport equation (CETE) \cite{suman2022novel}, extends the classical enstrophy framework by incorporating these compressibility effects, thereby enabling a detailed assessment of the relative contributions of vortex stretching \cite{lin2022physical}, dilatation, baroclinic torque, viscous transport, and viscous dissipation to the evolution of vortical structures. This formulation provides a physically consistent framework for investigating the mechanisms underlying the development of three-dimensionality in transonic compressible flows.

Motivated by these considerations, the present work investigates the early onset of transonic SBLI over the SHM1 airfoil using high-fidelity, 3D numerical simulations. Three operating conditions reported by Fujino {\it et al.} \citep{fujino2003natural} are considered, representing three distinct aerodynamic regimes: the initial development of a normal shock, drag divergence, and shock-induced separation. These cases are not treated as a parameter sweep; rather, they provide three representative operating regimes through which the early development of the interaction and its 3D characteristics can be examined. To identify the onset of 3D effects on the evolving shock system and surface aerodynamic response, 2D and 3D simulations are systematically compared. Subsequently, the CETE is employed to quantify the mechanisms governing the redistribution of enstrophy during the development of the shock-wave boundary layer interaction. Particular emphasis is placed on identifying how the relative contributions of vortex stretching, dilatation, baroclinic torque, and viscous transport evolve with increasing interaction strength, thereby providing a mechanistic explanation for the onset of three-dimensionality in transonic SBLI.

The remainder of the paper is organized as follows. Section 2 presents the governing equations, computational setup, and numerical methodology, including the selected operating conditions and validation of the numerical framework. Section 3 discusses the early transient evolution of the SBLI through the flow topology, surface aerodynamic response, 3D enstrophy evolution, and compressible enstrophy transport. The principal findings and concluding remarks are summarized in Sec. 4.

\section{Problem formulation of 3D transonic SBLI on SHM1 airfoil}

\begin{figure}[!h]
\centering
\includegraphics[width=.75\textwidth]{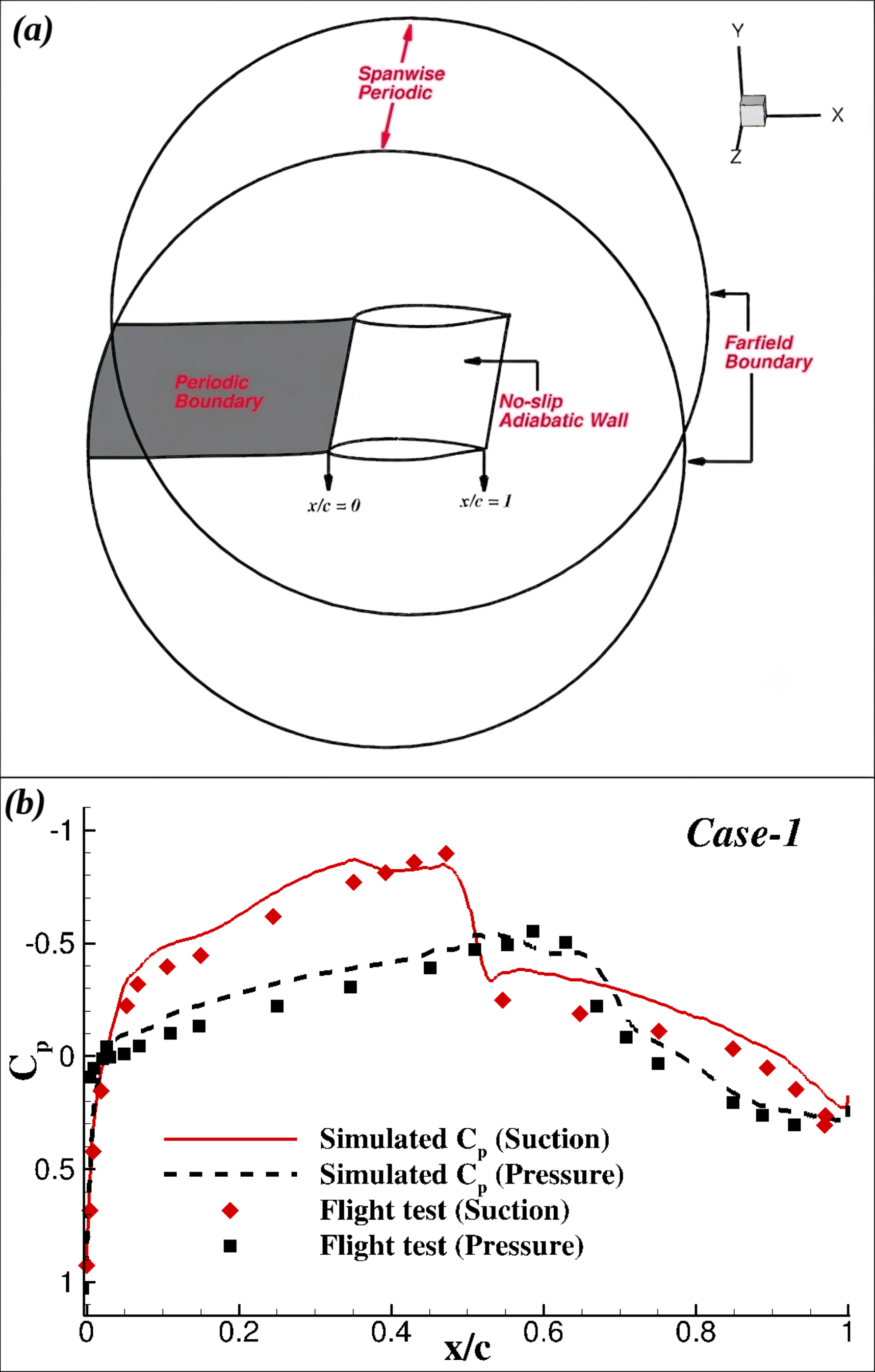}
\caption{(a) Schematic of the 3D computational domain and boundary conditions used for the high-fidelity simulations. \textbf{Note:} Schematic is not to scale; spanwise extent is enlarged and farfield distance is reduced for visual clarity. (b) Comparison of simulated time-averaged 2D $C_p$-distribution of Case-1 with Fujino's flight test data \citep{fujino2003natural}.}
\label{fig1}
\end{figure}

\noindent The SHM1 NLF airfoil developed by Honda R\&D and reported by Fujino {\it et al.} \cite{fujino2003natural} is adopted as the test configuration in the present study. This airfoil has been extensively characterized through flight tests and transonic wind-tunnel experiments, providing high-quality pressure distributions and aerodynamic data over a range of transonic operating conditions. Figure \ref{fig1}(a) illustrates the 3D computational domain and the associated boundary conditions. An O-type structured grid is generated using the hyperbolic grid generation algorithm in Pointwise. The mesh consists of 1251, 401, and 201 grid points in the azimuthal ($\xi$), wall-normal ($\eta$), and spanwise ($z$) directions, respectively, for all the cases reported here. The streamwise and spanwise grid spacings are uniform, while the wall-normal spacing is stretched using a hyperbolic-tangent distribution. Based on the local friction velocity, the corresponding near-wall resolutions are $\Delta x^+\approx1.25$-$3.70$, $\Delta y_{\min}^+\approx0.09$-$0.25$, and $\Delta z^+\approx 3.61$-$8.016$ for the operating conditions considered. The wall-normal mesh distribution, including a detailed view of the near-wall spacing, was presented previously in \cite{sengupta2021thermal}, where the same 2D mesh was employed. In addition, the mesh-independence study and validation of the 2D ($\xi,\eta$)-plane for the SHM1 airfoil were reported in \cite{sundaram2023non, sundaram2022non}. The corresponding 2D mesh is extruded uniformly in the spanwise direction to construct the present 3D computational domain. To facilitate structured grid generation, a cut is introduced along the $\xi$-direction, as illustrated in Fig. \ref{fig1}(a). The two faces created by this cut are treated as periodic boundaries, thereby preserving the continuity of all flow variables across the interface. The outer boundary is located at a radial distance of $16c$ from the airfoil surface, which minimizes the influence of the far-field boundary on the shock and wake dynamics while maintaining nearly uniform grid spacing in the wake region. The spanwise extent of the computational domain is set to $0.2c$, with periodic boundary conditions imposed in the spanwise direction, an approach adopted in prior high-fidelity simulations of transonic flows \cite{memmolo2018scrutiny}. The present spanwise extent is selected to permit the development of spanwise nonuniformity during the early evolution of the SBLI while maintaining a computationally tractable domain size. Previous studies have demonstrated that the predicted transonic flow can exhibit sensitivity to the spanwise domain width, particularly in regions associated with shock interaction and flow separation \cite{lusher2024effect}. The present study is restricted to the early transient evolution of the SBLI and does not seek to establish spanwise-domain independence or to characterize long-time transonic SBLI or buffet dynamics. Accordingly, the $0.2c$ span is used as a finite periodic domain that permits the development of 3D structures during the simulated interval. The mesh is clustered near the airfoil surface to resolve the viscous boundary layer, and additional grid refinement is applied in the streamwise direction near the leading edge, trailing edge, and the anticipated shock-induced separation region on the suction surface. This mesh distribution provides adequate resolution of the shock wave, boundary-layer development, and the ensuing SBLI \cite{sengupta2024thermal, sengupta2021thermal}. 

The numerical simulations are performed by solving the unsteady 3D compressible Navier-Stokes equations (NSE) \cite{hoffmann2000computational, pal2026direct, kumar2026supervised} formulated as:

\begin{equation}
\frac{\partial \hat{Q}}{\partial t^*}+\frac{\partial \hat{E_c}}{\partial x^*}+\frac{\partial \hat{F_c}}{\partial y^*}+\frac{\partial \hat{G_c}}{\partial z^*}=\frac{\partial \hat{E_v}}{\partial x^*}+\frac{\partial \hat{F_v}}{\partial y^*}+\frac{\partial \hat{G_v}}{\partial z^*},
\label{ge1}
\end{equation}

\noindent for the conserved variables given as
\begin{equation}
\hat{Q} = [ \rho^*; \; \rho^* u^*; \; \rho^* v^*; \; \rho^* w^*; \; \rho^* e_t^* ]^T.
\label{ge2}
\end{equation}
	
\noindent The convective flux variables $\hat{E}_c$, $\hat{F}_c$ and $\hat{G}_c$ are given as

\begin{equation}
\hat{E}_c = [ \rho^* u^*; \; \rho^* u^{*2} + p^*; \; \rho^* u^* v^* ; \; \rho^* u^* w^*; \; (\rho^* e_t^* + p^*) u^*]^T,
\label{ge3}
\end{equation}
 
\begin{equation}
\hat{F}_c = [ \rho^* v^*; \; \rho^* u^* v^*; \; \rho^* v^{*2} + p^*; \; \rho^* v^* w^*; \; 
(\rho^* e_t^* + p^*) v^* ]^T,
\label{ge4}
\end{equation}

\begin{equation}
\hat{G}_c = [ \rho^* w^*; \; \rho^* u^* w^*; \; \rho^* v^* w^*; \; \rho^* w^{*2} + p^*; \; (\rho^* e_t^* + p^*) w^* ]^T,
\label{ge5}
\end{equation}
	
\noindent and the viscous flux vectors $\hat{E}_v$, $\hat{F}_v$ and $\hat{G}_v$ are given as:

\begin{equation}
\hat{E}_v = [ 0; \; \tau^*_{xx}; \; \tau^*_{xy}; \; \tau^*_{xz}; \; u^* \tau^*_{xx} + v^* \tau^*_{xy} + w^* \tau^*_{xz} - q^*_x ]^T,
\label{ge6}
\end{equation}

\begin{equation}
\hat{F}_v = [ 0; \; \tau^*_{yx}; \; \tau^*_{yy}; \; \tau^*_{yz}; \; u^* \tau^*_{yx} + v^* \tau^*_{yy} + w^* \tau^*_{yz} - q^*_y ]^T,
\label{ge7}
\end{equation}

\begin{equation}
\hat{G}_v = [ 0; \; \tau^*_{zx}; \; \tau^*_{zy}; \; \tau^*_{zz}; \; u^* \tau^*_{zx} + v^* \tau^*_{zy} + w^* \tau^*_{zz} - q^*_z ]^T.
\label{ge8}
\end{equation}
	
Flow variables $\rho^*$, $p^*$, $u^*$, $v^*$, $w^*$, $T^*$ and $e_t^*$ represent dimensional density, pressure, Cartesian components of velocity, the absolute temperature and specific internal energy, respectively. The stress tensor $\tau^*_{ij}$, for $i,j = 1$ to 3, is related to the rate of strains as,

\begin{equation}
\tau^*_{ij} = \tau^*_{ji} =  \mu^* \biggl( \frac{\partial v_{j}^*}{\partial x_{i}^*} +  \frac{\partial v_{i}^*}{\partial x_{j}^*} \biggr );
		\; \tau^*_{ij} = \biggl[ \mu^* \biggl (\frac{\partial v_{j}^*}{\partial x_{i}^*} +  \frac{\partial v_{i}^*}{\partial x_{j}^*} \biggr ) + \lambda^* \frac{\partial v_{i}^*}{\partial x_{i}^*}   \biggr ] \; {\rm for} \; i = j;
\label{stress}
\end{equation}
	
\noindent The specific heat capacity ($C_v$) and thermal conductivity ($\kappa$) are constants in this formulation. The heat conduction terms $q^*_i$ are given as: $q^*_{i} = - \kappa \frac{\partial T^*}{\partial x^*_{i}}$. Sutherland's law is used for evaluating the dynamic viscosity ($\mu^*$) as a function of temperature ($T^*$), and Stokes' hypothesis is implemented for the bulk viscosity calculation. Additionally, the ideal gas relation, $p^*=\rho^* R^*T^*$ is the gas equation of state, which defines specific energy, $e_t^*$. With summation convention of Einstein, it is formulated as
	
\begin{equation}
e_t^* =\frac{p^*}{\rho^*(\gamma-1)} + \frac{v_{i}^*v_{i}^*}{2}.
\label{ge13}
\end{equation}
	
\noindent The independent and dependent variables are nondimensionalized \cite{hoffmann2000computational} as follows,
	
\begin{equation*}
x_i = \frac{x^*_{i}}{L}, \; v_i = \frac{v^*_{i}}{U_s}, \;t = \frac{t^*U_s}{L}, \; \rho = \frac{\rho^*}{\rho_s}, \; p = \frac{p^*}{p_s},\; e_t = \frac{e^*_t}{U^2_s}, \; T = \frac{T^*}{T_s}, \; \mu = \frac{\mu^*}{\mu_s}
\end{equation*}
 
\noindent where $L$, $U_s$, $T_s$ and $\rho_s$ are the length, velocity, temperature, and density scales and $p_s = \rho_sR^*T_s$, is the characteristic pressure obtained with the reference temperature and density. Non-dimensionalization is performed using temperature and density scales, specifically $T_s = 288.15K$ and $\rho_s = 1.2256 kg/m^3$ and $L$ is taken as the chord of the SHM1 airfoil. The time-scale is defined as $t = L/U_s$.

On the airfoil surface, shown in Fig. \ref{fig1}(a), no-slip boundary conditions are imposed, such that all velocity components satisfy $u=v=w=0$. The computational domain is assumed to be periodic in the spanwise direction over a width of $0.2c$. Thermal boundary conditions are specified as adiabatic, enforcing zero normal heat flux at the wall, i.e., $(q_x)_{wall} = (q_y)_{wall} = (q_z)_{wall} = 0$. At the outer boundary, the far-field flow is treated as locally one-dimensional in the $\eta$-direction, and a characteristic-based non-reflecting boundary condition \cite{pulliam1986implicit, sundaram2021multiscale} is employed. The incoming and outgoing characteristic variables are determined from the eigenstructure of the linearized one-dimensional Euler equations, with the formulation adapting to the local subsonic or supersonic flow conditions. At the inflow boundary, the incoming characteristic variables are prescribed using the free-stream conditions corresponding to the specified freestream Mach number ($M_{\infty}$), whereas at the outflow boundary, the flow variables are extrapolated from the interior whenever the local flow is supersonic. To minimize spurious reflections of acoustic disturbances and preserve the fidelity of the computed SBLI, the far-field boundary is located approximately $16c$ from the airfoil. This placement ensures that outgoing waves are sufficiently attenuated before reaching the computational boundary, thereby improving the effectiveness of the non-reflecting boundary treatment.

\subsection{Numerical methodology}

The governing equations are discretized using a high-order numerical framework based on dispersion-relation-preserving compact finite-difference schemes, enabling accurate resolution of broadband wave propagation and vortical structures in compressible flows \citep{sagaut2023global, sengupta2017global}. The convective flux derivatives are evaluated using the optimized upwind compact scheme (OUCS3) with explicit boundary closures, while the viscous terms are discretized using a second-order central difference (CD2) scheme to preserve the isotropic nature of the diffusive operators \citep{sagaut2023global}. Parallelization is achieved using the non-overlapping domain decomposition strategy proposed by Sundaram \textit{et al.} \citep{sundaram2023non}, which eliminates overlap regions at subdomain interfaces and retains the interior compact discretization across processor boundaries through global spectral optimization, thereby minimizing parallelization-induced errors. Time advancement is performed using the classical four-stage, fourth-order Runge--Kutta (RK4) scheme with a non-dimensional time step of $2.5 \times 10^{-6}$, ensuring adequate resolution of the relevant temporal scales. To suppress unresolved high-frequency numerical oscillations while preserving the resolved flow structures, a one-dimensional compact filter with a filter coefficient of 0.495 is applied independently along the streamwise, wall-normal, and spanwise directions at every time step. The numerical methodology has been validated in previous studies \citep{sundaram2023non,sengupta2025numerical} through comparisons of surface pressure distributions over the SHM1 airfoil with the flight-test measurements of Fujino \textit{et al.} \citep{fujino2003natural} at $Re_{\infty}=13.6\times10^{6}$, $M_{\infty}=0.62$, and $\alpha=0.27^{\circ}$, and at $Re_{\infty}=16.2\times10^{6}$, $M_{\infty}=0.72$, and $\alpha=0.38^{\circ}$, respectively. For completeness, the latter comparison is reproduced in Fig. \ref{fig1}(b), showing a good match with Fujino {\it et al.}'s flight test data \cite{fujino2003natural}. This provides a reference for the fidelity of the 2D numerical formulation at one of the operating conditions considered in the present study. The same numerical framework has also been validated against transonic wind-tunnel measurements for the NACA0012 airfoil \citep{harris1981two}, for which the large-scale facility substantially reduced wall-interference and 3D confinement effects commonly encountered in conventional transonic wind tunnels \citep{sengupta2013direct}. Since the present 3D simulations are restricted to the early evolution of the SBLI, they are not intended to provide statistically converged or time-averaged surface-pressure distributions. Consequently, a corresponding spanwise-averaged 3D $C_p$ distribution for comparison with the flight-test data is not presented.

Three operating conditions, summarized in Table \ref{tab1}, are considered to examine the early 3D evolution of transonic SBLI over the SHM1 airfoil. The freestream conditions are selected from the experimental and flight-test database of Fujino \textit{et al.} \citep{fujino2003natural}, thereby providing experimentally documented operating conditions for the present simulations. Rather than constituting a systematic variation of a single governing parameter, the three cases represent distinct transonic flow regimes characterized by different shock-interaction topologies. Case-1 corresponds to $M_{\infty}=0.72$, $Re_{\infty}=16.2\times10^{6}$, and $\alpha=0.38^{\circ}$, for which a predominantly normal shock develops on the suction surface \citep{sengupta2021thermal,chakraborty2022controlling}. Case-2 corresponds to $M_{\infty}=0.73$, $Re_{\infty}=8\times10^{6}$, and $\alpha=0.50^{\circ}$, representing the drag-divergence condition reported by Fujino \textit{et al.} \citep{fujino2003natural}, with a characteristic $\lambda$-shock structure developing over the suction surface. Case-3 corresponds to $M_{\infty}=0.78$, $Re_{\infty}=8\times10^{6}$, and $\alpha=0.50^{\circ}$, for which the shock interaction is accompanied by shock-induced boundary-layer separation and a pronounced $\lambda$-shock structure \citep{sengupta2024thermal}. These three operating conditions consequently provide distinct flow topologies against which the early departure of the 3D solution from corresponding 2D evolution can be examined. The simulations are restricted to the early onset of transonic SBLI computed till $t \le 5$ and are therefore not intended to characterize fully developed SBLI or long-time buffet dynamics. It is also important to distinguish the present operating conditions from the buffet regime identified experimentally by Fujino \textit{et al.} \citep{fujino2003natural}. In their experiments, an oscillating shock associated with the reported buffet limit was observed at $M_{\infty}=0.80$, $Re_{\infty} = 8\times10^6$, and $\alpha = 0.50^{\circ}$, whereas the present cases correspond to normal-shock development, drag divergence, and shock-induced separation, respectively.

\begin{table}[h!]
\centering
\caption{Numerical parameters used and description of the test cases reported.}
\vspace{1mm}
\begin{tabular}{|c| c| c| c| c|}
\hline
 Case & $M_{\infty}$ & $Re_{\infty}$ & $\alpha$  \\ [1ex]
 \hline\hline 
Case-1: Transonic normal shock appearance & 0.720 & $16.2 \times 10^{6}$ & $0.38^{\circ}$  \\ [0.5ex] 
Case-2: Drag-divergence condition & 0.730 & $8 \times 10^{6}$  & $0.50^{\circ}$ \\ [0.5ex] 
Case-3: Shock-induced separation regime & 0.780 & $8 \times 10^{6}$  & $0.50^{\circ}$  \\ [1ex]
 \hline
\end{tabular}

\label{tab1}
\end{table} 

\section{Results and discussion}
This section examines the early evolution of transonic SBLI over the SHM1 airfoil for the three operating regimes considered in the present study. The analysis first compares the 3D and corresponding 2D flow evolutions to characterize the development of the shock structure and associated flow topology. The temporal evolution of the surface pressure and skin-friction responses is then examined to identify when the developing 3D flow produces measurable departures from the corresponding 2D aerodynamic response. The development of the 3D vortical field is subsequently quantified through the volume-integrated streamwise and wall-normal enstrophies, providing a complementary global measure of the evolution of 3D vorticity. Finally, the compressible enstrophy transport equation (CETE) is employed to examine the relative contributions of vortex stretching, dilatation, baroclinic effects \cite{sengupta2023effects, sengupta2023multi}, and viscous transport to the evolving enstrophy budget.

\subsection{Evolving flow field during onset of transonic SBLI over NLF airfoil}

\begin{figure}[!h]
\centering
\includegraphics[width=0.9\textwidth]{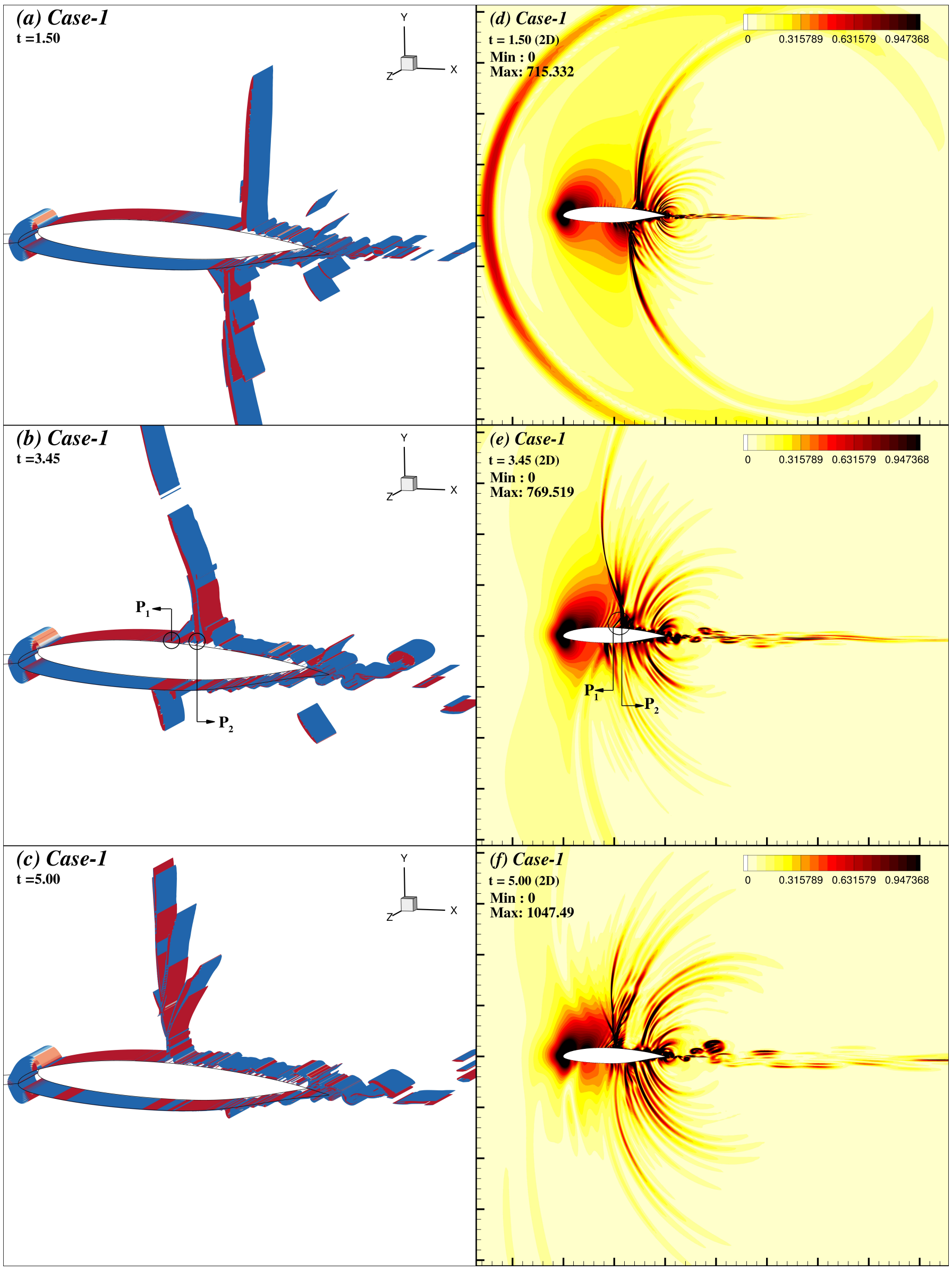}
\caption{Comparison of temporal flow evolution for Case-1: (a-c) 3D density gradient iso-surfaces, and (d-f) 2D density gradient magnitude contours at three non-dimensional timestamps ($t = 1.50, 3.45,$ and $5.00$). Annotations $P_1$ and $P_2$ in panels (b) and (e) highlight upstream and downstream positions on suction surface used to characterize the SBLI, respectively.}\label{fig2}
\end{figure}

\begin{figure}[!h]
\centering
\includegraphics[width=0.75\textwidth]{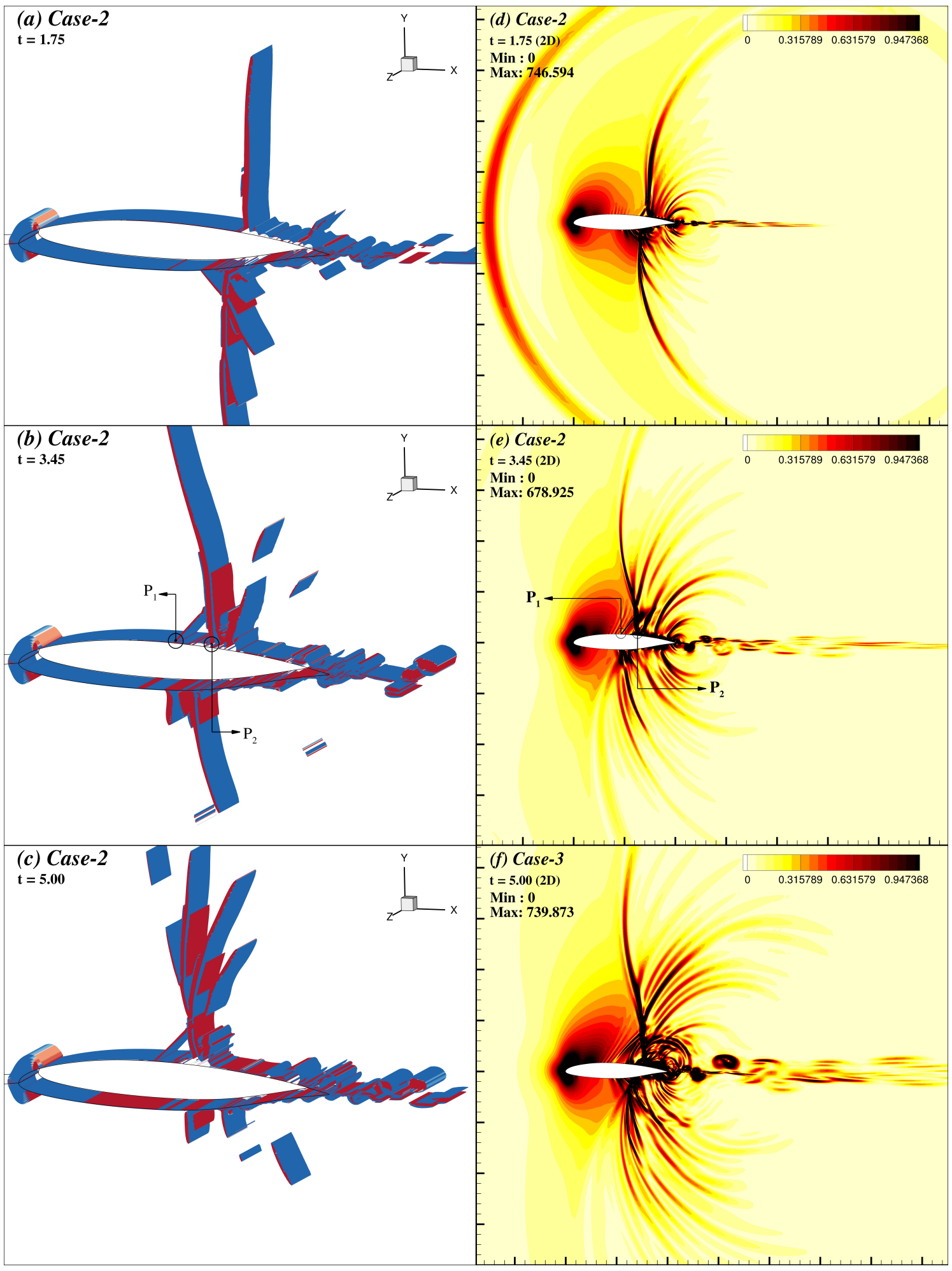}
\caption{Comparison of temporal flow evolution for Case-2: (a–c) 3D density gradient iso-surfaces, and (d–f) 2D density gradient magnitude contours at three non-dimensional timestamps ($t = 1.75, 3.45,$ and $5.00$), illustrating the $\lambda$-shock at drag divergence.}
\label{fig3}
\end{figure}

\begin{figure}[!h]
\centering
\includegraphics[width=0.75\textwidth]{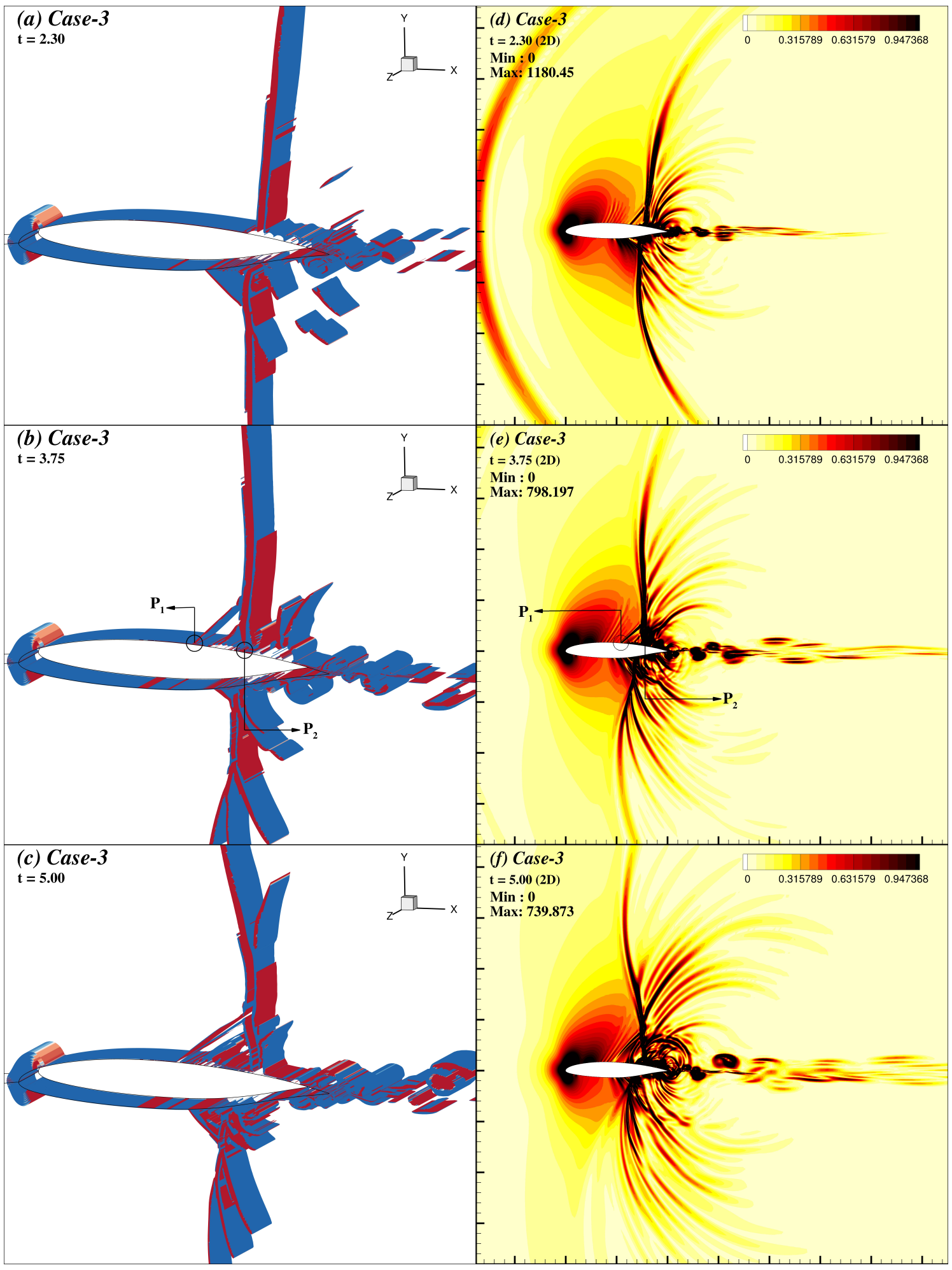}
\caption{Comparison of temporal flow evolution for Case-3: (a–c) 3D density gradient iso-surfaces, and (d–f) 2D density gradient magnitude contours at three non-dimensional timestamps ($t = 2.30, 3.75,$ and $5.00$), illustrating the shock-induced separation phenomenon.}
\label{fig4}
\end{figure}

Figure \ref{fig2} presents the evolution of the transonic SBLI over the SHM1 airfoil for Case-1 at three representative instants. The 3D density-gradient iso-surfaces, rendered using numerical shadowgraph visualization, are shown in Figs. \ref{fig2}(a-c), while the corresponding 2D density-gradient contours are presented in Figs. \ref{fig2}(d-f). The 2D results for transonic SBLI over the SHM1 airfoil were previously reported in \cite{sengupta2025numerical} and are used here as the 2D reference for comparison with the present 3D simulations. Figures \ref{fig2}-\ref{fig4} are intended primarily to establish the qualitative flow topology associated with the three operating regimes considered in Table \ref{tab1} and to provide a basis for comparison of the early 2D and 3D flow fields.

For Case-1, a well-defined shock develops on the suction surface and remains predominantly normal to the local flow direction \cite{tijdeman1977investigations} throughout the simulated interval. The locations \(P_1\) and \(P_2\) denote the upstream and downstream positions used to characterize the shock interaction, respectively. The corresponding 2D contours \cite{sengupta2021thermal} show a coherent and nearly planar shock structure, while the 3D visualization reveals the development of spanwise non-uniformity in the flow downstream of the shock. At subsequent times, these 3D  variations become increasingly apparent in the wake and in the downstream flow structures, although the primary shock remains spatially coherent. The Kutta waves \citep{lee1994role} and the wake exhibit appreciable spanwise variation in the 3D flow field, whereas the corresponding 2D flow field retains a planar structure. The overall shock topology, however, remains similar between the 2D and 3D solutions for the early onset of SBLI, and no large-scale shock-induced departure is observed during the simulated interval. Case-1 therefore provides the flow-topological baseline for the subsequent comparison with the drag-divergence and shock-induced-separation regimes.


Figure \ref{fig3} illustrates the temporal evolution of the transonic SBLI for Case-2, corresponding to the drag-divergence condition reported for the SHM-1 airfoil. The 3D density-gradient iso-surfaces in Figs. \ref{fig3}(a)-(c) are compared with the corresponding 2D density-gradient magnitude contours \cite{sengupta2025numerical} in Figs. \ref{fig3}(d)-(f) at \(t^*=1.75\), 3.45, and 5.00, respectively. A well-defined \(\lambda\)-shock structure \cite{sabnis2023review} develops over the suction surface, with the two characteristic shock-interaction locations denoted by \(P_1\) and \(P_2\). The overall \(\lambda\)-shock topology remains identifiable throughout the simulated interval, while the 3D flow field exhibits increasing spanwise nonuniformity in the SBLI and downstream regions that is absent from the corresponding 2D representation. At \(t=3.45\), the \(\lambda\)-shock structure and its interaction with the boundary layer are particularly evident in Fig. \ref{fig3}(e), while the corresponding 3D visualization in Fig. \ref{fig3}(b) reveals spanwise variations in the SBLI and downstream flow structures. At \(t=5.00\), these 3D variations become more pronounced, although the principal shock structure remains coherent in the spanwise direction. The 2D contours continue to capture the principal shock structure but, by construction, do not represent the spanwise evolution of the associated flow structures. Compared to Case-1 shown in Fig. \ref{fig2}, there is a noticeable downstream displacement of the shock structure and a broader extent of the SBLI over the suction surface.

Comparison with Fig. \ref{fig2} reveals that the downstream Kutta waves become increasingly complex at drag divergence, with stronger wave interactions developing in the wake. While the shock retains its overall spanwise coherence, the 3D density-gradient iso-surfaces in Figs. \ref{fig3}(a)-(c) exhibit more pronounced spanwise undulations than those observed in Case-1, indicating an increasing departure from a purely 2D flow field, consistent with previous investigations of transonic SBLI \cite{petrocchi2025numerical, bruce2010experimental}. The authors showed that increased SBLI strength enhances the adverse pressure gradient, intensifies the interaction region, and promotes increasingly complex unsteady flow behaviour prior to the onset of fully developed transonic SBLI. High-fidelity simulations \cite{garnier2010large} demonstrated that stronger transonic SBLI are accompanied by increasingly energetic downstream flow structures, while subsequent implicit large eddy simulations \cite{memmolo2018scrutiny} further highlighted the growing complexity of the wake and shear-layer dynamics as the interaction intensifies. The flow topology observed for this condition is consistent with the drag-divergence regime identified experimentally for the SHM1 airfoil by Fujino {\it et al. } \cite{fujino2003natural} and with the previously reported 2D numerical results \cite{sengupta2025numerical}.

Figure \ref{fig4} presents the temporal evolution of the transonic SBLI for Case-3, corresponding to the shock-induced-separation regime considered in the present study. The 3D density-gradient iso-surfaces in Figs. \ref{fig4}(a)-(c) are compared with the corresponding 2D density-gradient magnitude contours \cite{sengupta2024thermal} in Figs. \ref{fig4}(d)-(f) at \(t =2.30\), 3.75, and 5.00, respectively. A pronounced \(\lambda\)-shock structure develops over the suction surface, with the upstream and downstream shock-interaction locations denoted by \(P_1\) and \(P_2\), respectively. The appearance of the \(\lambda\)-structure is associated with shock-induced separation, for which the upstream compression waves generated by the separated flow interact with the primary shock structure, producing the characteristic two-leg shock configuration \cite{sabnis2023review}. The present flow topology is consistent with the experimental observations of Fujino {\it et al.} \cite{fujino2003natural}, wherein shock-induced separation is reported for the SHM-1 airfoil at \(M_\infty=0.78\).

Comparison with Figs. \ref{fig2} and \ref{fig3} shows a progressive change in the SBLI topology across the three operating conditions. For Case-1 shown in Fig. \ref{fig2}, the flow is dominated by a predominantly normal shock, whereas Case-2 at drag divergence exhibits a distinct \(\lambda\)-shock structure in Fig. \ref{fig3}. Case-3, on the other hand, shows that the \(\lambda\)-shock is accompanied by a separated region, producing a substantially more extended and spatially complex SBLI downstream of the shock. This evolution is consistent with the established dependence of shock-induced separation on the adverse pressure gradient imposed on the boundary layer \cite{sabnis2023review, babinsky2011shock}. The 3D flow field in Fig. \ref{fig4} further reveals a pronounced reorganization of the flow downstream of the separated region. At \(t=2.30\), the principal shock structure remains relatively coherent, while 3D structures are already evident downstream of the SBLI. With time, these structures develop into increasingly complex spanwise distributions and extend into the wake, as seen in Figs. \ref{fig4}(b) and \ref{fig4}(c). The corresponding 2D flow fields retain the principal shock structure and the shock-induced separation topology, but without the spanwise modulation observed in the 3D solution. This alters the extent and locations of the transonic SBLI along the suction surface, compared to the 3D flow field.  Thus, relative to the cases in Figs. \ref{fig2} and \ref{fig3}, the shock-induced-separation regime exhibits the most pronounced spatial development of the 3D downstream flow within the simulated interval.

Table \ref{tab2} compares the non-dimensional chordwise locations of the two upstream and downstream SBLI locations, \(P_1\) and \(P_2\), identified in the 2D and 3D flow fields presented in Figs. \ref{fig2}-\ref{fig4}.The comparison provides a quantitative measure of how closely the shock topology observed in the 3D simulations follows the corresponding 2D evolution across the three operating regimes.

\begin{table}[htbp]
\caption{Comparison of the non-dimensional chordwise locations of the upstream ($P_1$) and downstream ($P_2$) shock-interaction locations obtained from the 3D and corresponding 2D simulations for the three operating regimes shown in Figs. 2--4.}\label{tab2}%

\centering
\parbox{\linewidth}{
  \centering
  \setlength{\tabcolsep}{18pt} 
  \begin{tabular}{ccccc}
  \toprule
  \textbf{Case} & \multicolumn{2}{c}{\textbf{3D Flow}} & \multicolumn{2}{c}{\textbf{2D Flow}} \\
  \cmidrule(lr){2-3} \cmidrule(lr){4-5}
  & $P_1$ & $P_2$ & $P_1$ & $P_2$ \\
  \midrule
  Case-1 & 0.48 & 0.58 & 0.49 & 0.58 \\
  Case-2 & 0.51 & 0.62 & 0.47 & 0.63 \\
  Case-3 & 0.56 & 0.72 & 0.54 & 0.78 \\
  \bottomrule
  \end{tabular}
}
\end{table}

The locations of both \(P_1\) and \(P_2\) shift downstream across the selected operating conditions, with the change accompanied by a transition from a predominantly normal-shock configuration in Case-1, to a \(\lambda\)-shock structure at the drag-divergence condition in Case-2, and finally to a separated \(\lambda\)-shock configuration for Case-3. This progression is consistent with the experimentally identified flow regimes of the SHM-1 airfoil reported by Fujino {\it et al.} \cite{fujino2003natural}. From the SBLI perspective, an increase in the imposed shock pressure rise increases the adverse pressure gradient experienced by the boundary layer and, when separation occurs, the separated region modifies the pressure-wave system and extends the interaction region \cite{sabnis2023review}. 

For Case 1 (\(M_\infty=0.72\)), the 3D and 2D predictions are in close agreement. The \(P_1\) locations differ by only \(0.01c\), while \(P_2\) occurs at the same chordwise location, \(x/c=0.58\). This agreement is consistent with the predominantly coherent normal-shock topology observed in Fig. \ref{fig2}, indicating that the developing spanwise structures in the 3D solution have, at this stage, little influence on the identified shock-interaction locations. At the drag-divergence condition (\(M_\infty=0.73\)), the difference between the 2D and 3D solutions becomes more evident. The 3D value of \(P_1\) is located at \(x/c=0.51\), compared with \(x/c=0.47\) in the 2D solution, giving a displacement of \(0.04c\). In contrast, the \(P_2\) locations remain close, at \(x/c=0.62\) and \(0.63\), respectively. Thus, the principal difference at this condition is associated with the upstream portion of the \(\lambda\)-shock structure rather than with a systematic displacement of the entire shock structure. This observation is consistent with the more structured SBLI in Fig. \ref{fig3}, but the relatively small differences in \(P_2\) indicate that the overall downstream extent of the interaction remains similar between the 2D and 3D flow fields. The largest difference occurs for Case 3 (\(M_\infty=0.78\)), for which shock-induced separation accompanies the \(\lambda\)-shock structure. The \(P_1\) locations remain relatively close, with \(x/c=0.56\) for the 3D solution and \(x/c=0.54\) for the 2D solution. In contrast, the downstream location \(P_2\) occurs at \(x/c=0.72\) in the 3D solution and \(x/c=0.78\) in the 2D solution, corresponding to a difference of \(0.06c\). Thus, the principal difference in the separated regime for the 2D and 3D flow fields is associated with the downstream extent of the shock interaction. Such sensitivity of the SBLI extent to separation is consistent with established SBLI physics: once flow separation occurs, it generates additional compression waves and modifies the pressure field surrounding the main shock, thereby altering the overall SBLI topology \cite{sabnis2023review, delery2009some, tijdeman1977investigations}.

\subsection{Transient surface response during onset of transonic SBLI}

The transient evolution of the surface pressure and skin-friction distributions is examined through the pressure coefficient ($C_p$) and skin-friction coefficient ($C_{fx}$) to complement the qualitative flow visualizations. These quantities provide a quantitative description of the shock location, pressure recovery, and near-wall flow behaviour, thereby establishing the aerodynamic response associated with the evolving transonic SBLI.

\begin{figure}[!h]
\centering
\includegraphics[width=.65\textwidth]{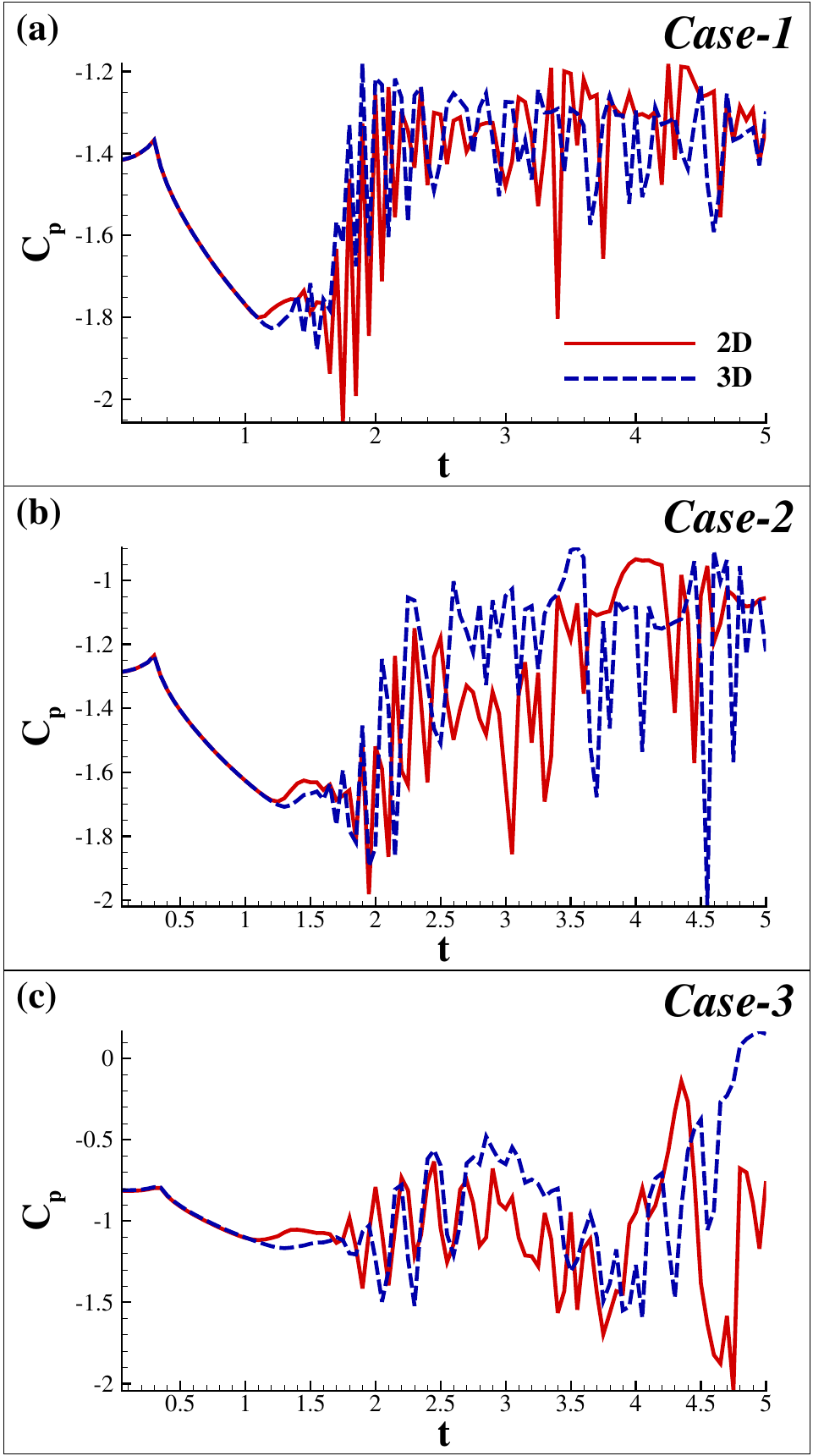}
\caption{Comparison of 2D and spanwise-averaged 3D pressure coefficient ($C_p$) time histories for (a) Case-1, (b) Case-2, and (c) Case-3. Pressure probes are positioned relative to corresponding mean shock location for each case.}
\label{fig5}
\end{figure}

To examine whether the developing 3D flow produces a measurable aerodynamic response, Fig. \ref{fig5} compares the temporal evolution of the surface pressure coefficient, \(C_p\), obtained from the 2D simulations \cite{sengupta2025numerical} with the corresponding spanwise-averaged 3D solutions. The pressure histories are extracted at the corresponding mean shock position for Cases 1-3 of Table \ref{tab1}, thereby allowing the temporal $C_p$ response to be compared at dynamically relevant locations despite the differences in shock position among the three operating conditions. The comparison is intended to identify the development of differences in the local $C_p$ response between the 2D and 3D evolutions, which enables the effects of 3D flow development to be distinguished from local spanwise variations.

The evolution of \(C_p\) exhibits two distinct stages. During the initial part of the simulations, the 2D and spanwise-averaged 3D pressure histories remain closely aligned for all three cases, indicating that the early surface-pressure response is initially governed predominantly by the corresponding 2D SBLI evolution. Subsequently, the pressure histories begin to diverge as 3D flow structures develop within and downstream of the interaction region. Such sensitivity of surface pressure to the evolving SBLI is consistent with experimental and numerical studies showing that the SBLI region is associated with enhanced wall-pressure fluctuations and that the pressure response is strongly affected by shock motion, boundary-layer separation, and the downstream flow organization \cite{mundell1986pressure, bernardini2011wall}.

For Case-1 shown in Fig. \ref{fig5}(a), the $C_p$ time-histories remain closely correlated over most of the simulated interval. Both the 2D and 3D solutions undergo a rapid change in pressure response around the development of the normal shock, followed by pronounced temporal fluctuations. Although individual excursions differ between the two flow fields, their overall range and temporal evolution remain comparable. This behaviour is consistent with the predominantly normal-shock topology observed in Fig. \ref{fig2} and the close correspondence of the \(P_1\) and \(P_2\) locations in Table \ref{tab2}. Within the limited early-transient interval considered here, the developing 3D flow produces only a modest modification of the selected local pressure response.

For Case-2 exhibiting drag-divergence, with appearance of a \(\lambda\)-shock as shown in Fig. \ref{fig3}, the correspondence between 2D and 3D flows becomes less persistent as shown in Fig. \ref{fig5}(b). The initial pressure evolution remains similar, but differences become apparent after approximately \(t\simeq2\), with the 3D solution exhibiting larger temporal excursions and less regular variation than the 2D solution. Such behaviour is physically consistent with the sensitivity of the SBLI to disturbances within the interaction region. The pressure disturbances associated with shock propagation and boundary-layer deformation can interact with the separated or strongly displaced boundary layer and modify the local wall-pressure response \cite{bernardini2011wall}. 

The distinction is more pronounced for Case-3 shown in Fig. \ref{fig5}(c), which demonstrates shock-induced separation. The 2D and 3D $C_p$ time-histories exhibit appreciable differences over much of the subsequent interval, including differences in the magnitude and timing of individual pressure excursions. The onset of this divergence occurs while the flow is still evolving and should therefore be interpreted as a signature of the developing 3D flow. The stronger sensitivity of the surface pressure in the separated case is consistent with established SBLI physics, in which separation modifies the pressure field through the coupled evolution of the shock, separated shear layer, and boundary layer \cite{delery2009some, sabnis2023review}.

It is important to emphasize that the \(C_p\) evolutions presented in Fig. \ref{fig5} are not statistically converged distributions for a fully developed or statistically stationary flow. The simulations are intentionally restricted to the early evolution, \(t\leq5\), and the purpose of Fig. \ref{fig5} is to compare the onset and early development of the 2D and 3D pressure responses under the three selected operating conditions. Consequently, the fluctuations should not be interpreted as statistically representative values of long-time pressure fluctuations, nor are they intended to characterize a fully developed turbulent state. This distinction is important because the shock and boundary-layer response in SBLI is intrinsically coupled to the evolving pressure field and flow structure, and substantial changes in local pressure can occur during the development of the interaction before a statistically stationary state is established \cite{dolling2001fifty}.

\begin{figure}[!h]
\centering
\includegraphics[width=.65\textwidth]{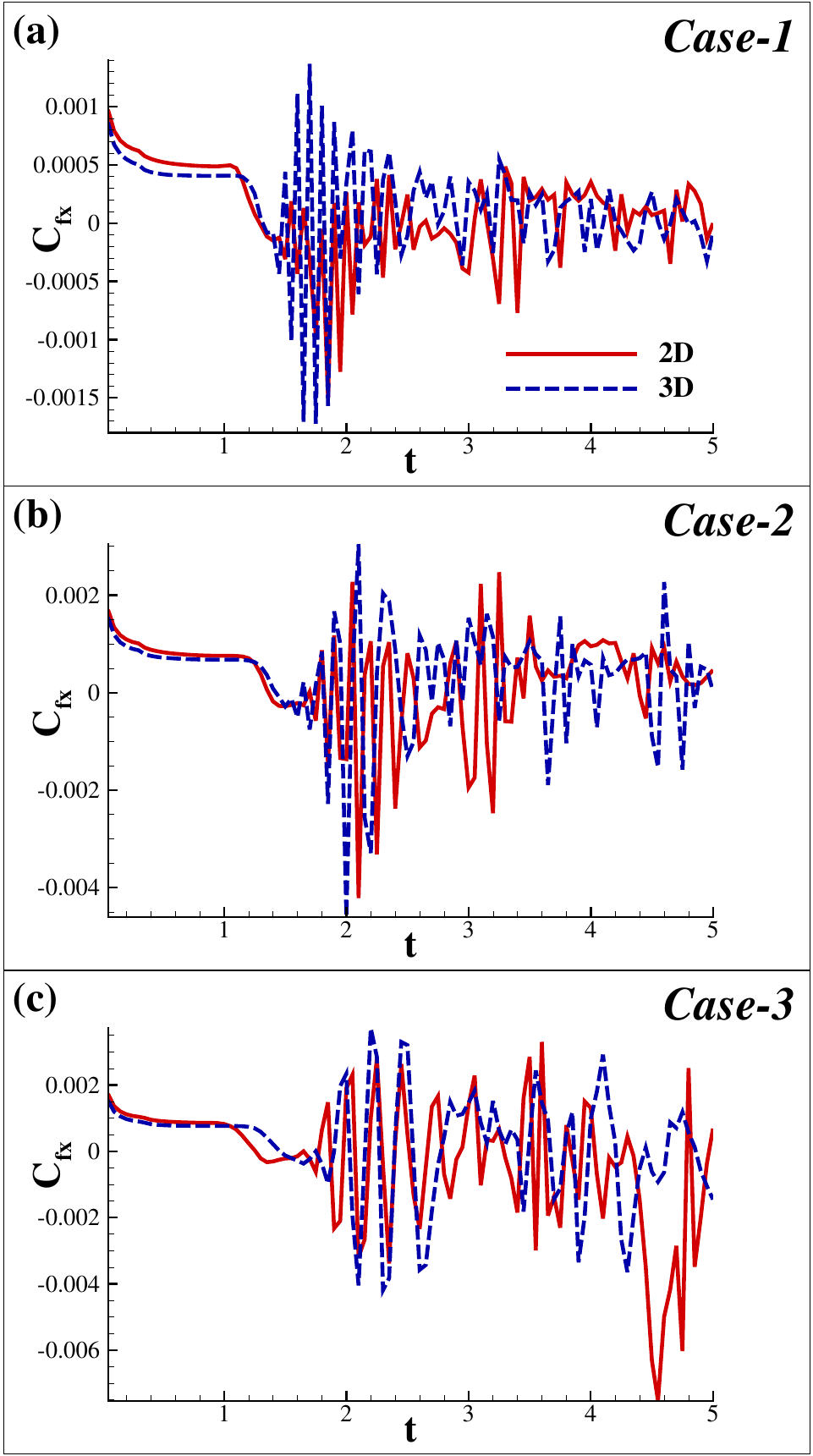}
\caption{Comparison of 2D and spanwise-averaged 3D skin friction coefficient ($C_{\mathrm{fx}}$) time histories for (a) Case-1, (b) Case-2, and (c) Case-3.}
\label{fig6}
\end{figure}

Figure \ref{fig6} compares the temporal evolution of streamwise skin-friction coefficient, \(C_{fx}\), obtained from 2D simulations \cite{sengupta2025numerical} with the corresponding spanwise-averaged 3D solutions. Unlike the $C_p$ considered in Fig. \ref{fig5}, which reflects the pressure response associated with the shock and the outer flow, \(C_{fx}\) directly measures the streamwise wall-shear response and is therefore particularly sensitive to the state of the boundary layer within the SBLI region. In particular, a reduction of \(C_{fx}\) toward zero indicates strong attenuation of wall shear, while a change in sign corresponds to a reversal of the local streamwise wall shear and is therefore indicative of instantaneous reverse flow at the monitored location. Skin friction is consequently a useful local indicator of the development of shock-induced separation \cite{liou1980interaction, swoboda1996shock, schulein2006skin}. Here, objective is to determine whether, and when, the developing 3D flow produces a different near-wall response from corresponding 2D evolution. This distinction is particularly relevant for SBLI because the adverse pressure gradient imposed by the shock progressively reduces the streamwise momentum of the boundary layer and can drive the wall shear toward zero and eventually reverse its direction as separation develops \cite{delery1985shock, sengupta2019direct, sundaram2020effects}.

For Case-1 shown in Fig. \ref{fig6}(a), the 2D and 3D $C_{fx}$ histories remain closely correlated during the initial evolution. Both solutions exhibit a gradual reduction in positive skin friction as the SBLI develops, followed by pronounced temporal excursions around \(t\approx1.5\)-2.0. The excursions include brief negative values in both solutions, but their subsequent evolution remains broadly comparable. Together with the close \(P_1\) and \(P_2\) locations in Table \ref{tab2} and the relatively coherent shock topology in Fig. \ref{fig2}, this indicates that the developing 3D flow has only a limited influence on the local wall-shear response during the simulated interval.

The response changes for Case-2 shown in Fig. \ref{fig6}(b). The initial reduction in \(C_{fx}\) remains similar in the 2D and 3D solutions, but their subsequent excursions become less correlated. Both solutions undergo repeated changes in the sign and magnitude of \(C_{fx}\), indicating intermittent local reversal of the streamwise wall shear at the monitored location. The 3D signal exhibits particularly large excursions around \(t\approx2\), followed by continued temporal variations that differ from those of the 2D solution. This behaviour is consistent with the structured \(\lambda\)-shock observed in Fig. \ref{fig3}. Importantly, however, the present \(C_{fx}\) history does not by itself establish a persistent separated region. Rather, it demonstrates that the near-wall response has become increasingly sensitive to the evolving SBLI.

The distinction between the two simulations is most evident for Case-3 shown in Fig. \ref{fig6}(c), for which shock-induced separation is observed in the flow field. The $C_{fx}$ histories exhibit frequent sign changes and substantially different excursions between the 2D and 3D solutions throughout the simulated interval. Such sign changes are consistent with the presence of local reverse flow associated with the separated or intermittently separated near-wall region. In transonic SBLI, the shock-induced adverse pressure gradient reduces wall shear and, once the boundary layer separates, the skin-friction response becomes closely linked to the separated shear layer and subsequent recovery of the wall flow \cite{schulein2006skin, delery1985shock}. Experimental work on transonic airfoils has likewise shown that the development and extent of local separation strongly influence the surface shear-stress distribution and the associated SBLI \cite{swoboda1996shock}.

\subsection{Evolution and transport of enstrophy during onset of transonic SBLI}

Enstrophy, defined as the squared magnitude of the vorticity vector, $\Omega = \vec{\omega} \cdot \vec{\omega}$, provides a measure of the rotational intensity of a flow. Unlike vorticity alone, enstrophy directly quantifies the strength of vortical structures and is therefore widely employed to investigate the generation, amplification, and dissipation of turbulence. In wall-bounded flows, regions of elevated enstrophy are closely associated with intense shear layers, coherent vortical structures, and the redistribution of kinetic energy through vortex stretching and viscous dissipation. Consequently, the transport of enstrophy provides valuable insight into the evolution of flow instabilities and the underlying mechanisms governing vorticity dynamics. In incompressible flows, the enstrophy transport equation \cite{sengupta2018enstrophy} has been extensively used to identify the competing roles of vortex stretching, viscous diffusion, and dissipation in the evolution of turbulent structures \cite{doering1995applied}.

For transonic SBLI, flow is inherently compressible, and the incompressible enstrophy framework is no longer sufficient. Strong density and pressure gradients generated across the shock wave introduce additional mechanisms for enstrophy production and redistribution that are absent in incompressible flows. In particular, dilatational effects, baroclinic torque \cite{sengupta2022thermally} arising from the misalignment of pressure and density gradients, and compressibility-modified viscous transport \cite{joshi2024highly, sengupta2023role} all contribute to the evolution of vorticity within the SBLI region. These mechanisms become particularly important in the vicinity of the shock and separated shear layer, where rapid thermodynamic changes strongly influence the local vorticity dynamics \cite{tijdeman1977investigations}. Consequently, a compressible enstrophy transport equation (CETE) \cite{suman2022novel} is required to isolate and quantify the individual physical processes governing the evolution of enstrophy in transonic SBLI.

\begin{figure}[!h]
\centering
\includegraphics[width=.6\textwidth]{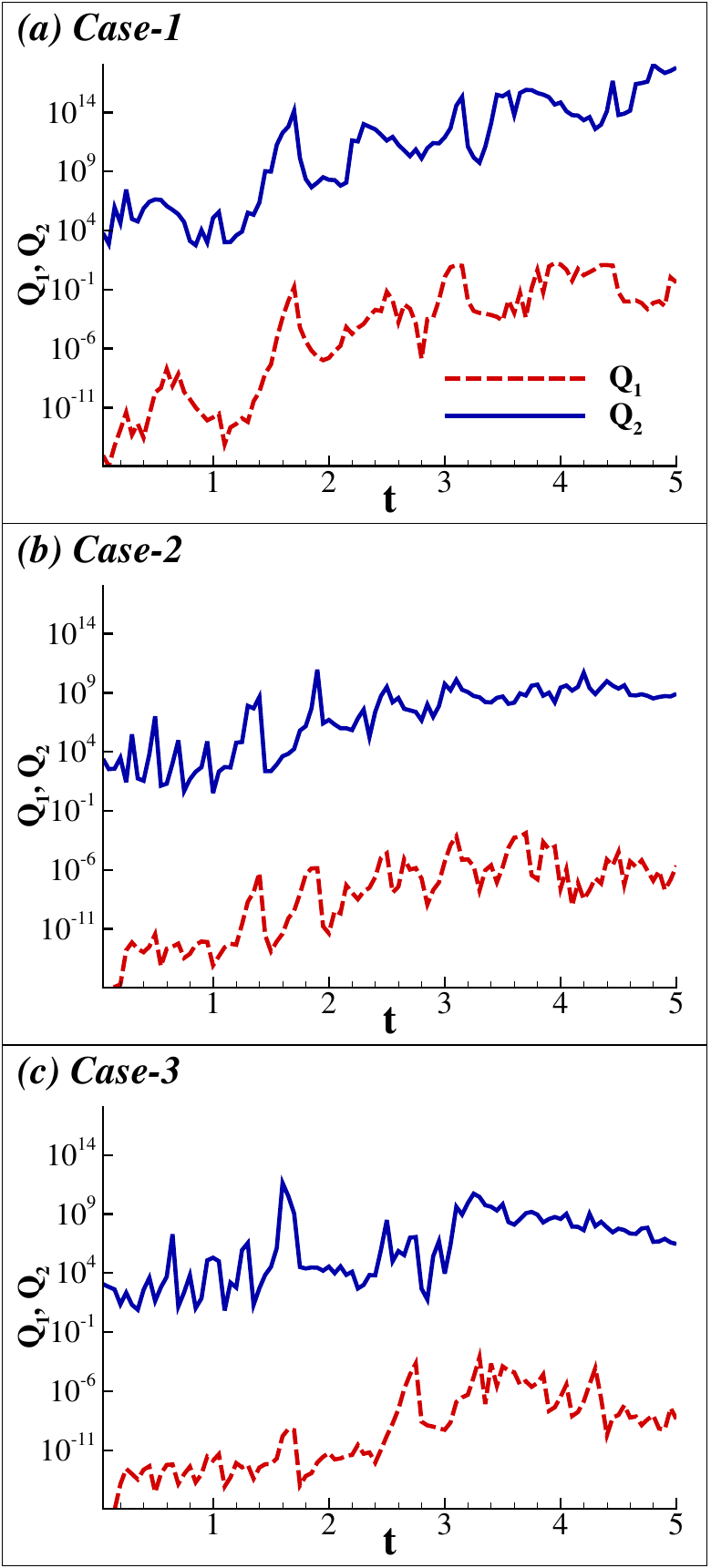}
\caption{Evolution of volume-integrated streamwise enstrophy $Q_1$,  and wall-normal enstrophy $Q_2$ for (a) Case-1, (b) Case-2 and (c) Case-3.}
\label{fig7}
\end{figure}

To characterize the development of three-dimensionality, the temporal evolution of the volume-integrated streamwise and wall-normal enstrophies is examined in Fig. \ref{fig7}. These quantities are defined as $Q_1=\int_V \omega_x^2\,dV,\; Q_2=\int_V \omega_y^2\,dV$, where \(\omega_x\) and \(\omega_y\) denote the streamwise and wall-normal components of vorticity, respectively. For a strictly two-dimensional flow, \(w=0\) and \(\partial/\partial z=0\), such that \(\omega_x=\omega_y=0\). Thus, the development of finite \(Q_1\) and \(Q_2\) provides a direct measure of the emergence and amplification of vorticity components associated with 3D flow. 

The temporal evolution in Fig. \ref{fig7} shows that both \(Q_1\) and \(Q_2\) remain finite and evolve through the simulated interval, with \(Q_2\) consistently exceeding \(Q_1\) for all three operating conditions. The two components exhibit pronounced temporal variations rather than a monotonic growth, indicating that the 3D vortical field evolves dynamically during the early evolution of SBLI. Case-1, corresponding to the appearance of a normal shock, exhibits the largest absolute values of both quantities, whereas Cases-2 and 3, corresponding to the drag-divergence and shock-induced-separation regimes, respectively, exhibit lower overall levels. These absolute magnitudes should not, however, be interpreted as a direct measure of the relative degree of three-dimensionality among the cases, since Case-1 is also characterized by a substantially higher \(Re_{\infty}\).

Of particular relevance is the relationship between Fig. \ref{fig7} and the surface-response histories presented in Figs. \ref{fig5} and \ref{fig6}. The departures of the 3D $C_p$ and $C_{fx}$ from their corresponding 2D solutions occur earlier than the pronounced changes observed in the volume-integrated \(Q_1\) and \(Q_2\). The \(C_p\) and \(C_{fx}\) quantities in Figs. \ref{fig5} and \ref{fig6} are evaluated locally within the SBLI region and are therefore sensitive to relatively small 3D disturbances in the vicinity of the shock and near-wall flow. In contrast, \(Q_1\) and \(Q_2\) quantify the integrated contribution of the corresponding vorticity components over the entire computational domain. Consequently, a localized three-dimensional disturbance can produce a measurable modification of the local shock or wall response before its contribution becomes sufficiently large or spatially widespread to produce a pronounced change in the volume-integrated enstrophy. Thus, the earlier departure of \(C_p\) and \(C_{fx}\) from the 2D solutions reflects the local aerodynamic sensitivity to developing 3D motion, whereas the subsequent evolution of \(Q_1\) and \(Q_2\) characterizes the spatial development and amplification of the associated 3D vortical field. 

This distinction is particularly evident in Case-3, where the development of shock-induced separation is accompanied by more pronounced temporal variation of \(Q_1\) and \(Q_2\) at \(t\sim2.5-3\). The corresponding surface-response histories in Figs. \ref{fig5} and \ref{fig6} also show a substantial departure between the 2D and 3D solutions during this stage. The concurrent changes in the surface response and 3D vorticity indicate that the development of separation is accompanied by enhanced 3D vortical dynamics within the SBLI region. Such changes in vortical structure and flow dynamics with interaction topology are consistent with the broader understanding of 3D SBLI reported in the literature \cite{gaitonde2023dynamics, sabnis2023review}.

The results establish the development of 3D vorticity but do not, by themselves, identify the physical mechanisms responsible for its redistribution during the evolving interaction. The CETE, derived from the compressible NSE by Suman {\it et al.} \cite{suman2022novel}, provides a physics-based framework for investigating vorticity evolution in compressible flows. Unlike its incompressible counterpart, the CETE explicitly incorporates compressibility-induced mechanisms \cite{joshi2023exploring, sengupta2025bifurcation} governing enstrophy transport. These include enstrophy production through velocity gradients and baroclinic torque generated by the misalignment of pressure and density gradients, modulation of enstrophy by flow dilatation, and the coupling between thermodynamic and vortical fields through viscous transport. Such processes are particularly relevant in transonic SBLI, where strong pressure and density gradients coexist with intense shear layers. The individual transport terms comprising the CETE \cite{suman2022novel}, are summarized below:

\begin{equation}
	\begin{aligned}
		\frac{D{\Omega }}{Dt}
		&= 2\vec{\omega } \cdot \left[(\vec{\omega} \cdot {\nabla}) \vec{V}\right] - 2({\nabla} \cdot \vec{V}) \Omega \\
		&\quad + \left(\frac{2}{\rho^{2}}\right) \vec{\omega } \cdot \left[\left({\nabla \rho} \times {{\nabla p}}\right)\right] -\left(\frac{2}{\rho^{2}}\right)\vec{\omega } \cdot \left[{\nabla \rho} \times {\nabla} \left(\lambda ({\nabla} \cdot \vec{V})\right)\right] \\
		&\quad +\left(\frac{4}{\rho}\right)\vec{\omega } \cdot \left[{\nabla} \times \left[{\nabla} \cdot \left(\mu S \right)\right]\right] -\left(\frac{4}{\rho^2}\right)\vec{\omega } \cdot \left({\nabla \rho} \times ({\nabla} \cdot \left(\mu S\right)) \right)
	\end{aligned}
	\label{CETE}
\end{equation}
	
The various terms of Eq. \eqref{CETE} are as follows:

\begin{itemize}
		
\item $2\vec{\omega } \cdot \left[(\vec{\omega} \cdot {\nabla}) \vec{V}\right]$ : Contribution to enstrophy due to vortex stretching (T1). 
		
\item $({\nabla} \cdot \vec{V}) \Omega$: Enstrophy growth/decay due to compressibility (T2). 
		
\item $\left(\frac{1}{\rho^{2}}\right) \vec{\omega } \cdot \left[\left({\nabla \rho} \times {{\nabla p}}\right)\right]$: Contribution from baroclinic term due to misalignment of gradients of pressure and density (T3). 
		
\item $\left(\frac{1}{\rho^{2}}\right)\vec{\omega } \cdot \left[{\nabla \rho} \times {\nabla} \left(\lambda ({\nabla} \cdot \vec{V})\right)\right]$ : Contribution due to misalignment of vorticity and bulk viscosity gradients (T4).
		
\item $\left(\frac{1}{\rho}\right)\vec{\omega } \cdot \left[{\nabla} \times \left[{\nabla} \cdot \left(\mu S \right)\right]\right]$: Diffusion of enstrophy due to viscous action (T5).
		
\item $\left(\frac{1}{\rho^2}\right)\vec{\omega } \cdot \left({\nabla \rho} \times ({\nabla} \cdot \left(\mu S\right)) \right)$: Contribution due to misalignment of gradients of density and divergence of viscous stresses (T6).
\end{itemize}

The vortex stretching term ($T_1$) governs the amplification of 3D vortical structures through the stretching of vortex lines and is most significant in separated shear layers and reattachment regions \cite{lin2022physical}. The dilatation terms account for enstrophy production or attenuation due to local compression and expansion and are unique to compressible flows \cite{sarkar1991analysis}. The baroclinic torque term represents vorticity generation arising from the misalignment of pressure and density gradients, making it particularly important in regions with strong thermodynamic gradients, such as shock waves, separation bubbles, and adverse pressure gradients \cite{sengupta2023compressibility, sengupta2024separation, sengupta2020effects}. Finally, the viscous transport and dissipation terms describe the redistribution and irreversible destruction of enstrophy through viscous effects, particularly at the smallest flow scales. 

\begin{figure}[!h]
\centering
\includegraphics[width=.9\textwidth]{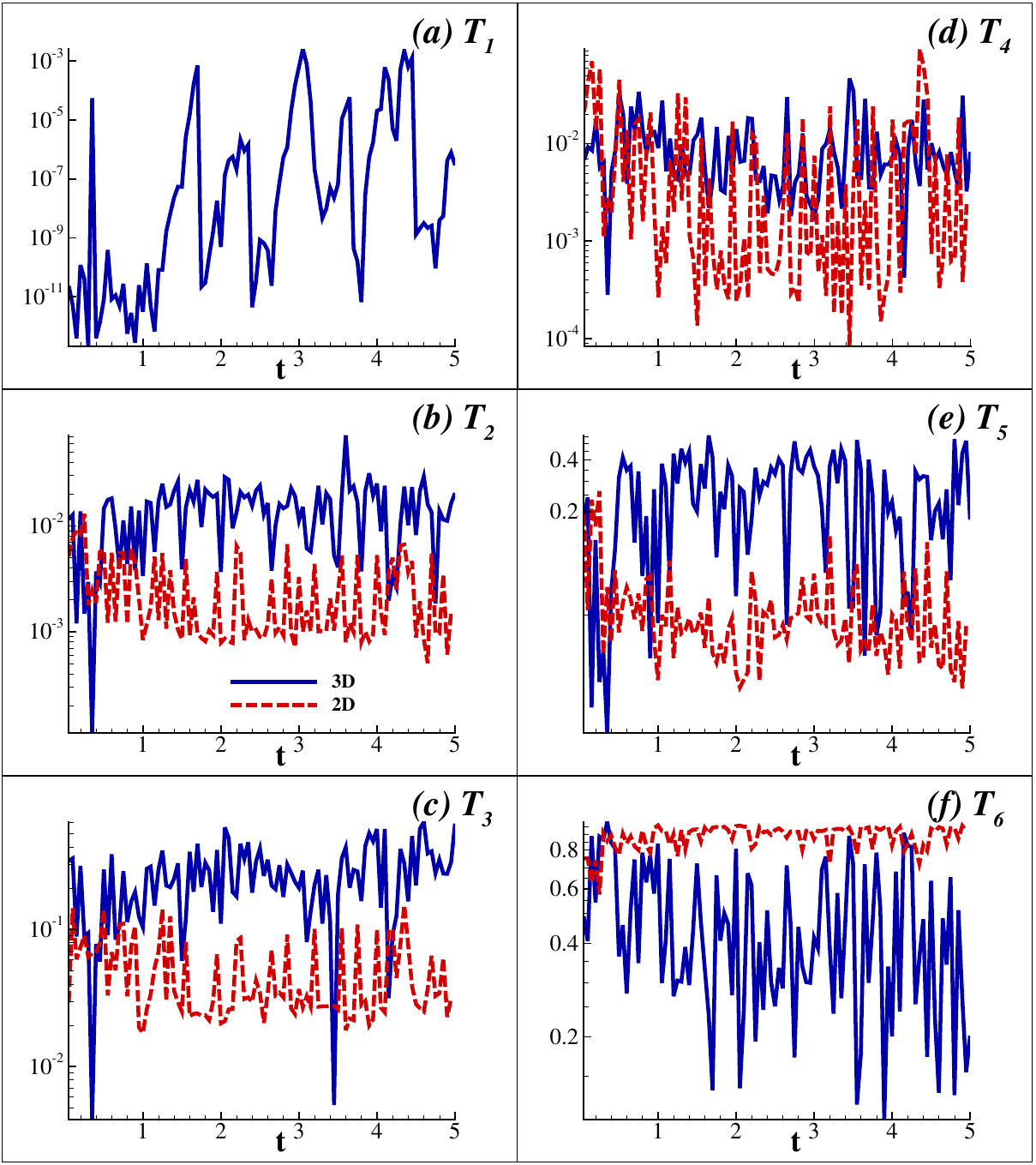}
\caption{Evolution of the fractional contribution of CETE budget terms, (a) $T_1$, (b) $T_2$, (c) $T_3$, (d) $T_4$, (e) $T_5$, and (f) $T_6$, comparing 2D and 3D flow configurations for Case-1.}
\label{fig8}
\end{figure}

Figure \ref{fig8} compares the temporal evolution of the normalized contributions of the six CETE budget terms obtained from 2D and spanwise-averaged 3D simulations for Case-1. The results provide a reference for examining how the development of three-dimensionality modifies the relative distribution of the enstrophy budget under the normal-shock regime. Since the quantities are presented as normalized contributions, the comparison is focused on relative importance and temporal variability of individual mechanisms rather than on their absolute dimensional magnitudes.

The most distinct feature of Fig. \ref{fig8} is the strong separation between the contributions associated with vortex stretching and those associated with dilatational, baroclinic, and viscous effects. The vortex-stretching term \(T_1\) in Fig. \ref{fig8}(a) remains several orders of magnitude smaller than the other contributions throughout the simulated interval. Although the 3D result exhibits intermittent increases in \(T_1\), particularly after \(t\approx1.5\), its magnitude remains substantially below the dominant terms. Thus, for the present normal-shock regime, vortex stretching does not constitute a leading contribution to the enstrophy budget during the early evolution. In contrast, the contributions of \(T_2\) and \(T_3\), shown in Figs. \ref{fig8}(b) and \ref{fig8}(c), are consistently higher in 3D flow than in corresponding 2D simulation over most of the interval. The increase in \(T_2\) indicates a greater relative contribution from dilatational effects when spanwise flow development is permitted, while the higher \(T_3\) contribution reflects a stronger relative role of baroclinic vorticity generation in 3D flow. The latter is particularly evident from the sustained separation between the 3D and 2D curves in Fig. \ref{fig8}(c). These differences indicate that the departure from the nominally 2D flow modifies the balance of compressibility-related mechanisms even though the overall shock topology remains predominantly normal, as observed in Fig. \ref{fig2}.

\begin{figure*}
\centering
\includegraphics[width=.9\textwidth]{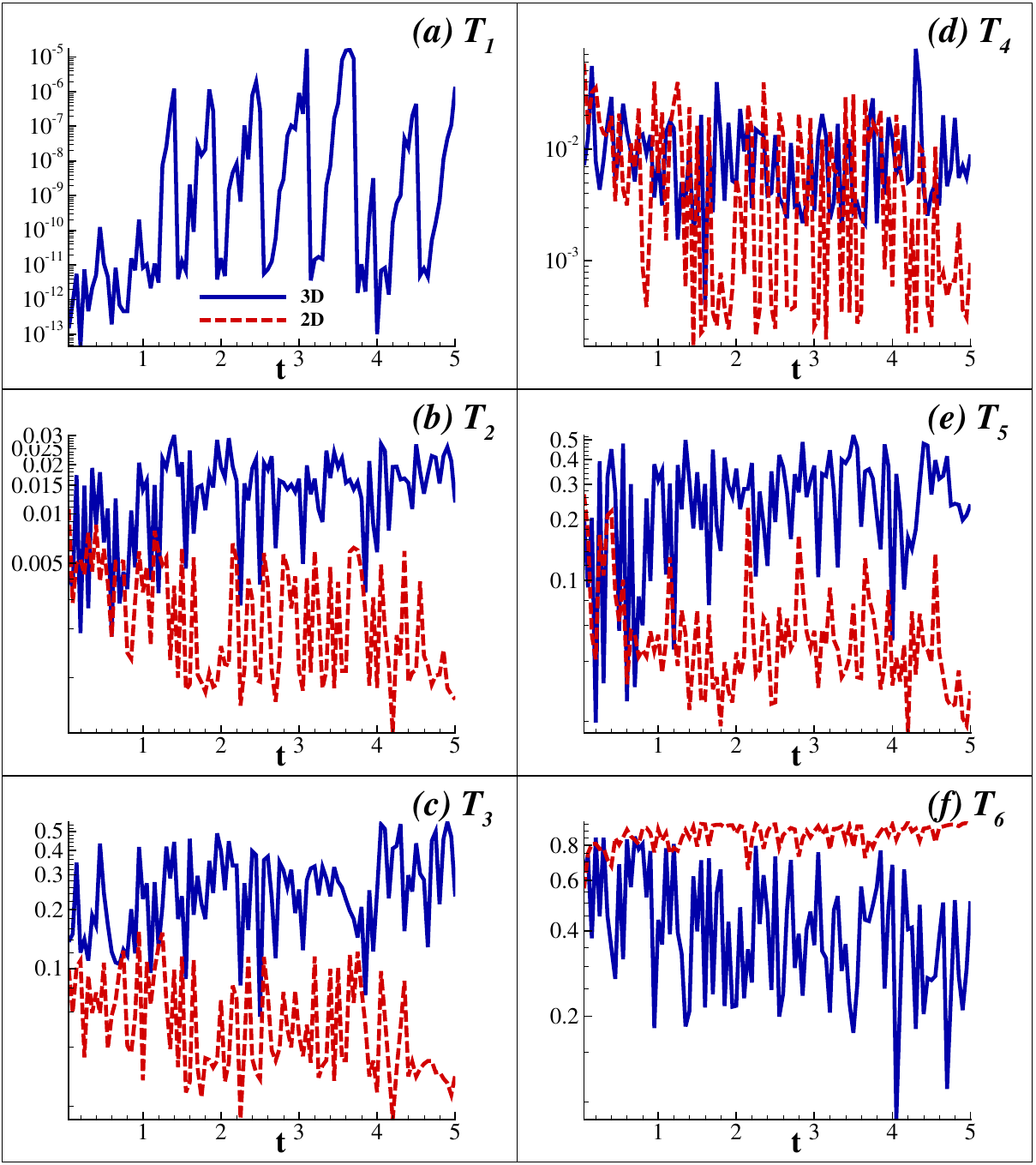}
\caption{Evolution of the fractional contribution of CETE budget terms, (a) $T_1$, (b) $T_2$, (c) $T_3$, (d) $T_4$, (e) $T_5$, and (f) $T_6$, comparing 2D and 3D flow configurations for Case-2.}
\label{fig9}
\end{figure*}

The viscous contributions exhibit an even more pronounced redistribution between the two flow configurations. The contribution of \(T_5\) in Fig. \ref{fig8}(e) is substantially larger in 3D flow, remaining in the range of approximately \(0.2{-}0.4\), whereas its 2D counterpart is generally much smaller. Conversely, \(T_6\) shown in Fig. \ref{fig8}(f) provides a dominant and comparatively persistent contribution in 2D simulation, with values close to \(0.8{-}0.9\), but becomes considerably more intermittent in 3D. The 3D \(T_6\) contribution fluctuates over a broad range during the entire interval, while the corresponding increase in \(T_5\) indicates a redistribution of the viscous contribution within the enstrophy budget. This behaviour is important because it shows that the influence of three-dimensionality is not limited to the appearance of additional vorticity components; it also alters how the existing enstrophy is redistributed among the individual transport and viscous mechanisms. In comparison, \(T_4\) shown in Fig. \ref{fig8}(d) remains of similar order in 2D and 3D simulations, although both exhibit substantial temporal fluctuations. No systematic separation between the two configurations is observed for this term, suggesting that its relative contribution is less sensitive to developing 3D structures under the present operating condition.

Figure \ref{fig9} presents the normalized CETE budget for Case-2, corresponding to the drag-divergence regime. In comparison with the normal-shock case in Fig. \ref{fig8}, the difference between the 2D and 3D enstrophy budgets becomes substantially more pronounced. The vortex-stretching term \(T_1\) shown in Fig. \ref{fig9}(a) remains several orders of magnitude smaller than the principal contributions, although 3D flow exhibits more pronounced intermittent excursions than those in Fig. \ref{fig8}. Thus, vortex stretching remains a secondary mechanism during the simulated interval. The most distinct changes occur in \(T_2\), \(T_3\), and \(T_5\) depicted in Figs. \ref{fig9}(b), \ref{fig9}(c) and \ref{fig9}(e), respectively. In the 3D simulation, \(T_2\) and \(T_3\) remain consistently above their 2D counterparts, with the separation between the two configurations particularly evident for the baroclinic contribution \(T_3\). Similarly, \(T_5\) is substantially larger in 3D and exhibits considerable temporal variability. These changes are accompanied by a marked redistribution of \(T_6\), as shown in Fig. \ref{fig9}(f). While \(T_6\) remains the dominant contribution in 2D, its 3D contribution is considerably reduced and becomes highly intermittent. In contrast, \(T_4\) shown in Fig. \ref{fig9}(d) remains of comparable order in the two configurations and does not exhibit a systematic demarcation between the 2D and 3D flows.

Relative to Case-1, the drag-divergence regime is characterized by a stronger redistribution of the normalized enstrophy budget between 2D and 3D solutions, particularly among the dilatational, baroclinic, and viscous terms. This redistribution occurs without a corresponding emergence of vortex stretching as a dominant mechanism. Thus, the developing 3D flow modifies the balance among the existing enstrophy-transport pathways even while the principal $\lambda$-shock remains identifiable.

\begin{figure}[!h]
\centering
\includegraphics[width=\textwidth]{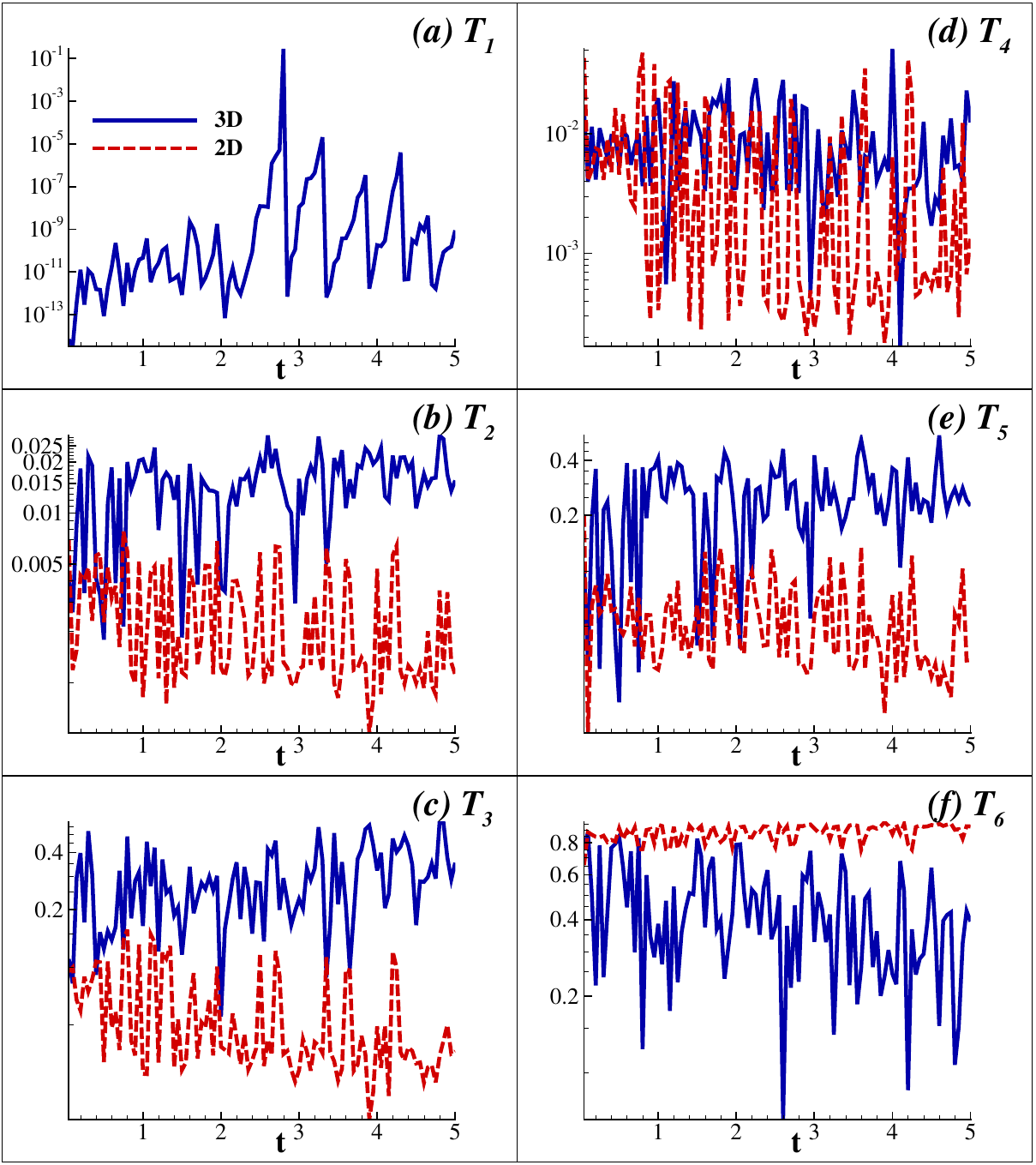}
\caption{Evolution of the fractional contribution of CETE budget terms, (a) $T_1$, (b) $T_2$, (c) $T_3$, (d) $T_4$, (e) $T_5$, and (f) $T_6$, comparing 2D and 3D flow configurations for Case-3.}
\label{fig10}
\end{figure}

Figure \ref{fig10} presents the normalized CETE budget for Case-3, corresponding to the shock-induced-separation regime. In comparison with Figs. \ref{fig8} and \ref{fig9}, the development of separation is accompanied by a more pronounced distinction between the 2D and 3D enstrophy budgets. The most evident differences occur in the dilatational, baroclinic, and viscous transport contributions. The 3D simulation exhibits consistently larger contributions from \(T_2\), \(T_3\), and \(T_5\) than the corresponding 2D simulation, as shown in Figs. \ref{fig10}(b), \ref{fig10}(c) and \ref{fig10}(e). A particularly clear separation between the two configurations is noted for \(T_3\) and \(T_5\). At the same time, \(T_6\) shown in Fig. \ref{fig10}(f), which remains relatively high and comparatively steady in the 2D solution, becomes strongly intermittent in the 3D simulation. Thus, as in Case-2, the principal effect of 3D flow development is a redistribution of the relative enstrophy budget among the existing transport mechanisms rather than the appearance of an entirely new dominant pathway.

A distinct feature of Case-3 is the behaviour of the vortex-stretching term \(T_1\), shown in Fig. \ref{fig10}(a). Unlike Cases-1 and 2, where \(T_1\) remains several orders of magnitude below the other contributions, the 3D solution in Case-3 exhibits intermittent excursions reaching substantially larger magnitudes, including a pronounced event around \(t\approx2.7\). Nevertheless, \(T_1\) is not the major contributor for most of the simulated interval. Its increased magnitude indicates that vortex stretching becomes an appreciable component of the evolving 3D enstrophy dynamics once shock-induced separation develops, rather than becoming the dominant mechanism governing the overall budget. The behaviour of \(T_1\) is consistent with the evolution identified in the preceding sections. In Case-3, the separated flow produces a more spatially developed 3D vortical field, reflected by the pronounced temporal variations in \(Q_1\) and \(Q_2\) in Fig. \ref{fig7} and the substantial departure of the local \(C_p\) and \(C_{fx}\) responses from their 2D counterparts. The CETE results indicate that this departure is accompanied not only by enhanced relative contributions from dilatational, baroclinic, and viscous transport, but also by the emergence of appreciable vortex stretching. The shock-induced separation regime therefore differs from Cases-1 and 2 primarily in the relative importance of the mechanisms within the evolving 3D enstrophy budget, rather than through a complete change in the set of mechanisms involved.

\section{Summary and conclusions}
The present study investigates the early evolution of three-dimensionality in transonic SBLI over the SHM1 NLF airfoil using high-fidelity 3D simulations at three operating conditions reported by Fujino {\it et al.} \cite{fujino2003natural}. The selected cases correspond to distinct flow regimes: development of a predominantly normal shock in Case-1, drag divergence with a \(\lambda\)-shock structure in Case-2, and shock-induced separation in Case-3. For each condition, the 3D solution is compared with the corresponding 2D simulation over the early interval \(t\leq5\). Surface pressure and skin friction are used to examine the local aerodynamic response, while the volume-integrated streamwise and wall-normal enstrophies, \(Q_1\) and \(Q_2\), quantify the development of the 3D vortical field. The compressible enstrophy transport equation (CETE) is subsequently used to identify the relative contributions of the underlying vorticity-transport mechanisms.

The flow visualizations show that the principal shock topology remains closely related to the corresponding 2D evolution across the three operating regimes, while the 3D simulations develop increasing spanwise nonuniformity in the SBLI region and downstream of it. The normal-shock in Case-1 exhibits comparatively limited modification of the primary shock topology, whereas the drag-divergence and shock-induced-separation regimes develop characteristic \(\lambda\)-shock structures accompanied by increasingly complex 3D downstream flow. The comparison of the SBLI extents along the suction surface further shows that the 3D and 2D solutions remain closely aligned in the normal-shock regime, while measurable differences appear in the drag-divergence and shock-induced separation configurations.

The surface-response histories provide a more sensitive indication of the developing 3D effects. The 3D and 2D \(C_p\) responses initially remain closely correlated, followed by measurable departures as the 3D flow develops. Similar departures are observed in \(C_{fx}\), with the separated case exhibiting the most pronounced differences in the near-wall response. The earlier response of these local quantities relative to the more pronounced changes in volume-integrated streamwise and wall-normal enstrophies (\(Q_1\) and \(Q_2\)) demonstrates an important distinction between local aerodynamic sensitivity and domain-integrated measures of 3D vorticity. A localized 3D disturbance within the shock or near-wall region can modify the surface response before its contribution becomes sufficiently large or spatially distributed to produce a comparable change in the volume-integrated enstrophy. Thus, the departure of the 3D solution from the 2D evolution is manifested first through local aerodynamic and near-wall quantities and subsequently through the spatial development of the global 3D vortical field.

The evolution of \(Q_1\) and \(Q_2\) confirms the development of non-zero streamwise and wall-normal vorticity components throughout the simulations, with \(Q_2\) consistently exceeding \(Q_1\) for all three operating conditions. Both quantities exhibit substantial temporal variability rather than monotonic growth, indicating that the 3D vortical field evolves dynamically during the early SBLI. The absolute levels of \(Q_1\) and \(Q_2\) differ among the cases and are not interpreted as a direct ranking of three-dimensionality, particularly because the operating conditions also differ in Reynolds number. Of particular significance is Case-3, where the temporal variation of both enstrophy measures becomes more pronounced around \(t\sim2.5{-}3\), coinciding with the development of shock-induced separation and the substantial departure of 3D surface response from corresponding 2D solution.

The CETE analysis provides the corresponding mechanistic interpretation. Across all three operating regimes, the principal enstrophy-transport mechanisms remain associated with viscous and compressibility-related processes, while the vortex-stretching contribution \(T_1\) is comparatively small during the normal-shock and drag-divergence regimes. The comparison of the 2D and 3D budgets shows that the development of three-dimensionality primarily modifies the relative distribution of the enstrophy budget rather than introducing a fundamentally different transport mechanism. In particular, 3D simulations exhibit larger relative contributions from the dilatational (\(T_2\)), baroclinic (\(T_3\)), and viscous transport (\(T_5\)) terms, accompanied by a redistribution of the contribution associated with viscous stress term, \(T_6\). The bulk-viscosity contribution \(T_4\) remains comparatively similar between the 2D and 3D configurations. A distinct change occurs when shock-induced separation develops. In Case-3, vortex stretching exhibits intermittent excursions that are substantially larger than those observed in the normal-shock and drag-divergence cases, although \(T_1\) does not become the dominant contribution to the overall normalized budget. The separated regime is therefore distinguished not by the appearance of a new enstrophy-transport mechanism, but by a change in the relative importance and temporal activity of the existing mechanisms, including a measurable contribution from vortex stretching. In conjunction with the evolution of \(Q_1\) and \(Q_2\), this result links the development of the separated 3D vortical field to a modified balance of enstrophy-transport processes.

Overall, the results demonstrate that the early departure of a transonic SBLI from its corresponding 2D evolution is best characterized through a combination of local surface response, global 3D vorticity measures, and enstrophy-transport analysis. The local \(C_p\) and \(C_{fx}\) responses identify the aerodynamic consequences of developing 3D motion, \(Q_1\) and \(Q_2\) quantify the subsequent evolution of the associated 3D vortical field, and the CETE analysis identifies the redistribution of the underlying transport mechanisms. Within the simulated early-time interval, the normal-shock and drag-divergence regimes are characterized primarily by redistribution among dilatational, baroclinic, and viscous pathways, whereas the development of shock-induced separation is accompanied by a more pronounced and intermittent contribution from vortex stretching. These findings provide a mechanistic description of the early 3D evolution of transonic SBLI over an NLF airfoil without invoking a fully developed or statistically stationary flow state. Future work will extend the present analysis on early onset of SBLI to longer-time evolution and systematically examine how the identified 3D vorticity mechanisms develop in fully developed turbulent SBLI.

\bmhead{Acknowledgements}
The authors would like to acknowledge the use of high-performance computing facility CHANDRASEKHAR at Indian Institute of Technology Dhanbad for all the computations reported here.

\section*{Declaration of Interests}
The authors report no conflict of interest. No funding was received to assist with the preparation of this manuscript.

\section*{Data Availability}
The datasets produced from the high-fidelity simulations are available from the corresponding author upon reasonable request. 

\bibliography{sbli_onset}

@article{hoffmann2000computational,
  author = {K. A. Hoffmann and S. T. Chiang},
  title = {Computational fluid dynamics volume I},
  journal = {Eng. Educ. Syst.},
  year = {2000}
}

@article{sagaut2023global,
  author = {P. Sagaut and V. K. Suman and P. Sundaram and M. K. Rajpoot and Y. G. Bhumkar and S. Sengupta and A. Sengupta and T. K. Sengupta},
  title = {{Global spectral analysis: Review of numerical methods}},
  journal = {Comput. Fluids},
  pages = {105915},
  year = {2023},
  publisher = {Elsevier}
}

@article{sundaram2021multiscale,
  author = {P. Sundaram and T. K. Sengupta and A. Sengupta and V. K. Suman},
  title = {{Multiscale instabilities of Magnus--Robins effect for compressible flow past rotating cylinder}},
  journal = {Phys. Fluids},
  volume = {33},
  number = {3},
  year = {2021},
  publisher = {AIP Publishing}
}

@article{pal2026direct,
  title={Direct numerical simulations of separation-induced transition in a low-pressure turbine under varying compressibility},
  author={Pal, Priya and Guha, Abhijeet and Sengupta, Aditi},
  journal={Physics of Fluids},
  volume={38},
  number={3},
  year={2026},
  publisher={AIP Publishing}
}

@article{sengupta2017global,
  title={{Global spectral analysis of multi-level time integration schemes: Numerical properties for error analysis}},
  author={Sengupta, Tapan K and Sengupta, Aditi and Saurabh, Kumar},
  journal={Applied Mathematics and Computation},
  volume={304},
  pages={41--57},
  year={2017},
  publisher={Elsevier}
}

@article{sengupta2023effects,
  title={{Effects of stabilizing and destabilizing thermal gradients on reversed shear-stratified flows: Combined Kelvin--Helmholtz Rayleigh--Taylor instability}},
  author={Sengupta, Aditi and Joshi, Bhavna},
  journal={Physics of Fluids},
  volume={35},
  number={1},
  year={2023},
  publisher={AIP Publishing}
}

@article{sengupta2023multi,
  title={{Multi-layer Rayleigh--Taylor instability: Consequences for naturally occurring stratified mixing layers}},
  author={Sengupta, Aditi and Ulloa, Hugo N and Joshi, Bhavna},
  journal={Physics of Fluids},
  volume={35},
  number={10},
  year={2023},
  publisher={AIP Publishing}
}

@article{joshi2023exploring,
  title={Exploring role of aspect ratio for compressible flow in a rectangular lid-driven cavity with a vertical temperature gradient},
  author={Joshi, Bhavna and Sengupta, Aditi and Sundaram, Prasannabalaji},
  journal={Physics of Fluids},
  volume={35},
  number={6},
  year={2023},
  publisher={AIP Publishing}
}

@article{sengupta2024separation,
  title={{Separation-induced transition on a T106A blade under low and elevated free stream turbulence}},
  author={Sengupta, Aditi and Gupta, Nivedita and Ubald, Bryn Noel},
  journal={Physics of Fluids},
  volume={36},
  number={2},
  year={2024},
  publisher={AIP Publishing}
}

@article{sengupta2025bifurcation,
  title={Bifurcation analysis of compressible flow past a rotating cylinder for high rotation rate},
  author={Sengupta, Aditi and Kumar, Santosh and Kumar, Sanjeev},
  journal={Physics of Fluids},
  volume={37},
  number={12},
  year={2025},
  publisher={AIP Publishing}
}

@article{joshi2024highly,
  title={{Highly resolved peta-scale direct numerical simulations: Onset of Kelvin--Helmholtz Rayleigh--Taylor instability via pressure pulses}},
  author={Joshi, Bhavna and Sengupta, Tapan K and Sundaram, Prasannabalaji and Sengupta, Aditi},
  journal={Computers \& Fluids},
  volume={284},
  pages={106442},
  year={2024},
  publisher={Elsevier}
}

@article{sengupta2023role,
  title={{Role of unstable thermal stratifications on the Rayleigh--Taylor instability}},
  author={Sengupta, Aditi and Verma, Atul Kumar},
  journal={Computers \& Fluids},
  volume={252},
  pages={105773},
  year={2023},
  publisher={Elsevier}
}

@article{sengupta2022thermally,
  title={{Thermally stratified free shear layers: Combined Kelvin--Helmholtz Rayleigh--Taylor instability}},
  author={Sengupta, Aditi and Joshi, Bhavna and Verma, Atul Kumar},
  journal={Physics of Fluids},
  volume={34},
  number={9},
  year={2022},
  publisher={AIP Publishing}
}

@article{sengupta2019direct,
  title={Direct numerical simulation of vortex-induced instability for a zero-pressure-gradient boundary layer},
  author={Sengupta, Aditi and Suman, VK and Sengupta, Tapan K},
  journal={Physical Review E},
  volume={100},
  number={3},
  pages={033118},
  year={2019},
  publisher={APS}
}

@article{sundaram2020effects,
  title={Effects of free stream excitation on the boundary layer over a semi-infinite flat plate},
  author={Sundaram, Prasannabalaji and Suman, Vajjala K and Sengupta, Aditi and Sengupta, Tapan K},
  journal={Physics of Fluids},
  volume={32},
  number={9},
  year={2020},
  publisher={AIP Publishing}
}

@article{sengupta2020effects,
  title={Effects of forced frequency oscillations and free stream turbulence on the separation-induced transition in pressure gradient dominated flows},
  author={Sengupta, Aditi and Tucker, Paul},
  journal={Physics of Fluids},
  volume={32},
  number={10},
  year={2020},
  publisher={AIP Publishing}
}

@article{kumar2026supervised,
  title={Supervised machine learning applied to bifurcation analysis of weakly compressible flow past a rotating cylinder},
  author={Kumar, Sanjeev and Kumar, Santosh and Sengupta, Aditi},
  journal={Computers \& Fluids},
  pages={107219},
  year={2026},
  publisher={Elsevier}
}

@article{pulliam1986implicit,
  author = {T. H. Pulliam},
  title = {Implicit solution methods in computational fluid dynamics},
  journal = {Appl. Numer. Math.},
  volume = {2},
  number = {6},
  pages = {441--474},
  year = {1986},
  publisher = {Elsevier}
}

@article{sundaram2023non,
  author = {P. Sundaram and A. Sengupta and V. K. Suman and T. K. Sengupta},
  title = {{Non-overlapping high-accuracy parallel closure for compact schemes: Application in multiphysics and complex geometry}},
  journal = {ACM Trans. Parallel Comput.},
  volume = {10},
  number = {1},
  pages = {1--28},
  year = {2023},
  publisher = {ACM}
}

@article{suman2022novel,
  author = {V. K. Suman and P. Sundaram and J. K. Puttam and A. Sengupta and T. K. Sengupta},
  title = {{A novel compressible enstrophy transport equation-based analysis of instability during Magnus--Robins effects for high rotation rates}},
  journal = {Phys. Fluids},
  volume = {34},
  number = {4},
  year = {2022},
  publisher = {AIP Publishing}
}

@article{memmolo2018scrutiny,
  title={Scrutiny of buffet mechanisms in transonic flow},
  author={Memmolo, Antonio and Bernardini, Matteo and Pirozzoli, Sergio},
  journal={International Journal of Numerical Methods for Heat \& Fluid Flow},
  volume={28},
  number={5},
  pages={1031--1046},
  year={2018},
  publisher={Emerald Publishing Limited}
}

@article{lusher2024effect,
  title={Effect of tripping and domain width on transonic buffet on periodic NASA-CRM airfoils},
  author={Lusher, David J and Sansica, Andrea and Hashimoto, Atsushi},
  journal={AIAA Journal},
  volume={62},
  number={11},
  pages={4411--4430},
  year={2024},
  publisher={American Institute of Aeronautics and Astronautics}
}

@article{fujino2003natural,
  title={Natural-laminar-flow airfoil development for a lightweight business jet},
  author={Fujino, M. and Yoshizaki, Y. and Kawamura, Y.},
  journal={Journal of Aircraft},
  volume={40},
  number={4},
  pages={609--615},
  year={2003}
}

@article{chakraborty2022controlling,
  title={Controlling transonic shock--boundary layer interactions over a natural laminar flow airfoil by vortical and thermal excitation},
  author={Chakraborty, Ardhendu and Roy, Arkadyuti Ghosh and Sundaram, Prasannabalaji and Sengupta, Aditi and Sengupta, Tapan K},
  journal={Physics of Fluids},
  volume={34},
  number={8},
  year={2022},
  publisher={AIP Publishing}
}

@article{sengupta2022comparative,
  title={Comparative study of transonic shock--boundary layer interactions due to surface heating and cooling on an airfoil},
  author={Sengupta, T. K. and Chakraborty, A. and Roy, A. G. and Sengupta, A. and Sundaram, P.},
  journal={Physics of Fluids},
  volume={34},
  number={4},
  year={2022},
  publisher={AIP Publishing}
}

@article{sengupta2024thermal,
  title={Thermal optimization of shock-induced separation in a natural laminar airfoil operating at off-design conditions},
  author={Sengupta, Aditi and Shandilya, Narottam},
  journal={Physics of Fluids},
  volume={36},
  number={4},
  year={2024},
  publisher={AIP Publishing}
}

@article{sengupta2021thermal,
  title={Thermal control of transonic shock-boundary layer interaction over a natural laminar flow airfoil},
  author={Sengupta, Tapan K and Roy, Arkadyuti Ghosh and Chakraborty, Ardhendu and Sengupta, Aditi and Sundaram, Prasannabalaji},
  journal={Physics of Fluids},
  volume={33},
  number={12},
  year={2021},
  publisher={AIP Publishing}
}

@article{sundaram2022non,
  title={{A non-overlapping high accuracy parallel subdomain closure for compact scheme: Onset of Rayleigh-Taylor instability by ultrasonic waves}},
  author={Sundaram, Prasannabalaji and Sengupta, Aditi and Sengupta, Tapan K},
  journal={Journal of Computational Physics},
  volume={470},
  pages={111593},
  year={2022},
  publisher={Elsevier}
}

@techreport{harris1981two,
  title={Two-dimensional aerodynamic characteristics of the NACA 0012 airfoil in the Langley 8 foot transonic pressure tunnel},
  author={Harris, Charles D},
  year={1981}
}

@article{sengupta2013direct,
  title={Direct numerical simulation of 2D transonic flows around airfoils},
  author={Sengupta, Tapan K and Bhole, Ashish and Sreejith, NA},
  journal={Computers \& Fluids},
  volume={88},
  pages={19--37},
  year={2013},
  publisher={Elsevier}
}

@article{tijdeman1977investigations,
  title={Investigations of the transonic flow around oscillating airfoils},
  author={Tijdeman, Hendrik},
  journal={NLR-TR 77090 U},
  year={1977},
  publisher={Nationaal Lucht-en Ruimtevaartlaboratorium}
}

@article{lee1994role,
  title={Role of Kutta waves on oscillatory shock motion on an airfoil},
  author={Lee, BHK and Murty, H and Jiang, H},
  journal={AIAA journal},
  volume={32},
  number={4},
  pages={789--796},
  year={1994}
}

@article{sengupta2025numerical,
  title={A numerical study of design and off-design operations of SHM1 airfoil},
  author={Sengupta, Aditi and Guha, Abhijeet},
  journal={Frontiers in Aerospace Engineering},
  volume={4},
  pages={1531916},
  year={2025},
  publisher={Frontiers Media SA}
}

@incollection{petrocchi2025numerical,
  title={Numerical Study of Unsteady Shock/Boundary Layer Interaction},
  author={Petrocchi, Andrea and Steijl, Rene and Barakos, George N},
  booktitle={Towards Effective Flow Control and Mitigation of Shock Effects in Aeronautical Applications},
  pages={193--220},
  year={2025},
  publisher={Springer}
}

@article{bruce2010experimental,
  title={An experimental study of transonic shock/boundary layer interactions subject to downstream pressure perturbations},
  author={Bruce, PJK and Babinsky, H},
  journal={Aerospace Science and Technology},
  volume={14},
  number={2},
  pages={134--142},
  year={2010},
  publisher={Elsevier}
}

@inproceedings{garnier2010large,
  title={Large-eddy simulation of transonic buffet over a supercritical airfoil},
  author={Garnier, E and Deck, S},
  booktitle={Direct and Large-Eddy Simulation VII: Proceedings of the Seventh International ERCOFTAC Workshop on Direct and Large-Eddy Simulation, held at the University of Trieste, September 8-10, 2008},
  pages={549--554},
  year={2010},
  organization={Springer}
}

@article{dolling2001fifty,
  title={Fifty years of shock-wave/boundary-layer interaction research: what next?},
  author={Dolling, David S},
  journal={AIAA journal},
  volume={39},
  number={8},
  pages={1517--1531},
  year={2001}
}

@book{babinsky2011shock,
  title={Shock wave-boundary-layer interactions},
  author={Babinsky, Holger and Harvey, John K},
  volume={32},
  year={2011},
  publisher={Cambridge University Press}
}

@article{delery2009some,
  title={Some physical aspects of shock wave/boundary layer interactions},
  author={D{\'e}lery, Jean and Dussauge, Jean-Paul},
  journal={Shock waves},
  volume={19},
  number={6},
  pages={453--468},
  year={2009},
  publisher={Springer}
}

@book{doering1995applied,
  title={{Applied analysis of the Navier-Stokes equations}},
  author={Doering, Charles R and Gibbon, John D},
  number={12},
  year={1995},
  publisher={Cambridge university press}
}

@article{sengupta2018enstrophy,
  title={An enstrophy-based linear and nonlinear receptivity theory},
  author={Sengupta, A. and Suman, V. K. and Sengupta, T. K. and Bhaumik, S.},
  journal={Physics of Fluids},
  volume={30},
  number={5},
  year={2018},
  publisher={AIP Publishing}
}

@article{sengupta2023compressibility,
  title={Compressibility effects on the flow past a T106A low-pressure turbine cascade},
  author={Sengupta, Aditi and Sundaram, Prasannabalaji},
  journal={Physics of Fluids},
  volume={35},
  number={10},
  year={2023},
  publisher={AIP Publishing}
}

@article{sarkar1991analysis,
  title={The analysis and modelling of dilatational terms in compressible turbulence},
  author={Sarkar, Sutanu and Erlebacher, Gordon and Hussaini, M Yousuff and Kreiss, Heinz Otto},
  journal={Journal of Fluid Mechanics},
  volume={227},
  pages={473--493},
  year={1991},
  publisher={Cambridge University Press}
}

@article{lin2022physical,
  title={Physical origin of vortex stretching and twisting: Viscous or inertial forces},
  author={Lin, LM and Wu, YX},
  journal={Physics of Fluids},
  volume={34},
  number={9},
  year={2022},
  publisher={AIP Publishing}
}

@article{krishnan2017review,
  title={Review of hybrid laminar flow control systems},
  author={Krishnan, KSG and Bertram, O and Seibel, O},
  journal={Progress in Aerospace Sciences},
  volume={93},
  pages={24--52},
  year={2017},
  publisher={Elsevier}
}

@article{karpuk2024investigation,
  title={Investigation of Hybrid Laminar Flow Control Capabilities from the Flight Envelope Perspective},
  author={Karpuk, Stanislav and Mosca, Valerio},
  journal={Journal of Aircraft},
  pages={1--26},
  year={2024},
  publisher={American Institute of Aeronautics and Astronautics}
}

@article{halila2019effects,
  title={Effects of boundary layer transition on the aerodynamic analysis of high-lift systems},
  author={Halila, Gustavo Luiz Olichevis and Antunes, Alexandre Pequeno and Da Silva, Ricardo Galdino and Azevedo, Jo{\~a}o Luiz F},
  journal={Aerospace Science and Technology},
  volume={90},
  pages={233--245},
  year={2019},
  publisher={Elsevier}
}

@book{cole2012transonic,
  title={Transonic aerodynamics},
  author={Cole, Julian D and Cook, L Pamela},
  year={2012},
  publisher={Elsevier}
}

@article{larsson2013reynolds,
  title={Reynolds-and Mach-number effects in canonical shock--turbulence interaction},
  author={Larsson, Johan and Bermejo-Moreno, Ivan and Lele, Sanjiva K},
  journal={Journal of Fluid Mechanics},
  volume={717},
  pages={293--321},
  year={2013},
  publisher={Cambridge University Press}
}

@inproceedings{gross2018numerical,
  title={Numerical analysis of laminar and turbulent shock-wave boundary layer interactions},
  author={Gross, Andreas and Lee, Sunyoung},
  booktitle={2018 Fluid Dynamics Conference},
  pages={4033},
  year={2018}
}

@article{quadros2018numerical,
  title={Numerical investigation of transitional shock-wave/boundary-layer interaction in supersonic regime},
  author={Quadros, Russell and Bernardini, Matteo},
  journal={AIAA Journal},
  volume={56},
  number={7},
  pages={2712--2724},
  year={2018},
  publisher={American Institute of Aeronautics and Astronautics}
}

@article{dussauge2008shock,
  title={Shock/boundary-layer interactions: Possible sources of unsteadiness},
  author={Dussauge, J-P and Piponniau, S},
  journal={Journal of Fluids and Structures},
  volume={24},
  number={8},
  pages={1166--1175},
  year={2008},
  publisher={Elsevier}
}

@inproceedings{toure2018numerical,
  title={{Numerical and experimental study of nominal 2-D shock-wave/turbulent boundary layer interactions}},
  author={Tour{\'e}, P. and Schuelein, E.},
  booktitle={2018 Fluid Dynamics Conference},
  pages={3395},
  year={2018}
}

@article{barre1996experimental,
  title={Experimental study of a normal shock/homogeneous turbulence interaction},
  author={Barre, S and Alem, D and Bonnet, JP},
  journal={AIAA journal},
  volume={34},
  number={5},
  pages={968--974},
  year={1996}
}

@article{ceci2025grid,
  title={Grid resolution requirements for DNS of shock/boundary-layer interactions},
  author={Ceci, Alessandro and Palumbo, Andrea and Pirozzoli, Sergio},
  journal={Computers \& Fluids},
  pages={106892},
  year={2025},
  publisher={Elsevier}
}

@article{sabnis2023review,
  title={A review of three-dimensional shock wave--boundary-layer interactions},
  author={Sabnis, Kshitij and Babinsky, Holger},
  journal={Progress in Aerospace Sciences},
  volume={143},
  pages={100953},
  year={2023},
  publisher={Elsevier}
}

@article{gaitonde2023dynamics,
  title={Dynamics of three-dimensional shock-wave/boundary-layer interactions},
  author={Gaitonde, Datta V and Adler, Michael C},
  journal={Annual Review of Fluid Mechanics},
  volume={55},
  number={1},
  pages={291--321},
  year={2023},
  publisher={Annual Reviews}
}

@article{zauner2019direct,
  title={Direct numerical simulations of transonic flow around an airfoil at moderate Reynolds numbers},
  author={Zauner, Markus and De Tullio, Nicola and Sandham, Neil D},
  journal={AIAA Journal},
  volume={57},
  number={2},
  pages={597--607},
  year={2019},
  publisher={American Institute of Aeronautics and Astronautics}
}

@article{bernardini2011wall,
  title={The wall pressure signature of transonic shock/boundary layer interaction},
  author={Bernardini, Matteo and Pirozzoli, Sergio and Grasso, Francesco},
  journal={Journal of Fluid Mechanics},
  volume={671},
  pages={288--312},
  year={2011},
  publisher={Cambridge University Press}
}

@article{mundell1986pressure,
  title={Pressure fluctuations caused by transonicshock/boundary-layer interaction},
  author={Mundell, ARG and Mabey, DG},
  journal={The Aeronautical Journal},
  volume={90},
  number={897},
  pages={274--282},
  year={1986},
  publisher={Cambridge University Press}
}

@article{liou1980interaction,
  title={Interaction between a normal shock wave and a turbulent boundary layer at high transonic speeds. Part II: Wall shear stress},
  author={Liou, Meng-Sing and Adamson Jr, Thomas Charles},
  journal={Zeitschrift f{\"u}r angewandte Mathematik und Physik ZAMP},
  volume={31},
  number={2},
  pages={227--246},
  year={1980},
  publisher={Springer}
}

@article{swoboda1996shock,
  title={Shock boundary-layer interaction on transonic airfoils for laminar and turbulent flow},
  author={Swoboda, Marius and Nitsche, Wolfgang},
  journal={Journal of aircraft},
  volume={33},
  number={1},
  pages={100--108},
  year={1996}
}

@article{schulein2006skin,
  title={Skin friction and heat flux measurements in shock/boundary layer interaction flows},
  author={Sch{\"u}lein, Erich},
  journal={AIAA journal},
  volume={44},
  number={8},
  pages={1732--1741},
  year={2006}
}

@article{delery1985shock,
  title={Shock wave/turbulent boundary layer interaction and its control},
  author={Delery, Jean M},
  journal={Progress in Aerospace Sciences},
  volume={22},
  number={4},
  pages={209--280},
  year={1985},
  publisher={Elsevier}
}

\end{document}